\documentclass[a4paper, onecolumn, 11pt, unpublished]{quantumarticle}
\pdfoutput=1
\usepackage[english]{babel}

\usepackage[letterpaper,top=2cm,bottom=2cm,left=1.5cm,right=1.5cm,marginparwidth=2cm]{geometry}

\newcommand\orcidicon[1]{\href{https://orcid.org/#1}{\mbox{\scalerel*{
\begin{tikzpicture}[yscale=-1,transform shape]
\pic{orcidlogo};
\end{tikzpicture}
}{|}}}}

\usepackage{amsmath,amssymb,amsfonts}
\usepackage[legacycolonsymbols]{mathtools}
\usepackage{mathrsfs}
\usepackage{amsfonts}
\usepackage{graphicx}
\usepackage{textcomp}
\usepackage[colorlinks=true, allcolors=blue]{hyperref}
\usepackage{physics}
\usepackage{caption}
\usepackage{url}
\usepackage{ulem}

\usepackage{subcaption}

\newcommand{\Hilbertspace}{\mathscr{H}}
\newcommand{\Linearmap}[2]{\mathscr{L}(#1, #2)}
\newcommand{\vectorisation}{\mathrm{vec}}
\newcommand{\standerdmatrixbasis}[1]{\hat{\Delta}_{#1}}
\newcommand{\Choimatrix}[1]{\Upsilon_{#1}}
\newcommand{\1}{\mbox{1}\hspace{-0.25em}\mbox{l}}
\newcommand{\doubleket}[1]{| #1 \rangle \! \rangle}
\newcommand{\doublebra}[1]{\langle \! \langle #1 |}
\newcommand{\doublebraket}[2]{\langle \! \langle #1 | #2 \rangle \! \rangle}
\newcommand{\timesteps}[1]{\mathbf{T}_{{#1}}}
\newcommand{\outcomes}[1]{\mathbf{x}_{{#1}}}
\newcommand{\CPmaps}[1]{\mathscr{A}_{{#1}}}
\newcommand{\code}{\mathcal{C}}
\newcommand{\decoder}{\mathcal{D}}
\newcommand{\recovery}{\mathcal{R}}
\newcommand{\stabilizers}{\mathcal{G}}
\newcommand{\logicals}{\mathcal{L}}
\newcommand{\Pauligroup}[1]{\mathscr{P}_{#1}}
\newcommand{\projector}{\pi}
\newcommand{\choiprojector}{\mathbf{\Pi}}
\newcommand{\probability}[1]{\mathbb{P}\left(#1\right)}
\newcommand{\argmin}[1]{\underset{#1} {\operatorname{argmin}}\:}
\newcommand{\argmax}[1]{\underset{#1} {\operatorname{argmax}}\:}

\title{Tensor-network decoders for process tensor descriptions of non-Markovian noise}

\author{Fumiyoshi Kobayashi}
\affiliation{Graduate School of Engineering Science, Osaka University\\
1-3 Machikaneyama, Toyonaka, Osaka 560-8531, Japan}
\affiliation{Center for Quantum Information and Quantum Biology\\
Institute for Open and Transdisciplinary Research Initiatives, Osaka University\\ 1-2 Machikaneyama, Toyonaka, Osaka 560-0043, Japan}

\author{Hidetaka Manabe}
\affiliation{Graduate School of Engineering Science, Osaka University\\
1-3 Machikaneyama, Toyonaka, Osaka 560-8531, Japan}

\author{Gregory A. L. White}
\affiliation{Dahlem Center for Complex Quantum Systems, Freie Universität Berlin, Germany}

\author{Terry Farrelly}
\affiliation{ARC Centre of Excellence in Engineered Quantum Systems and\\ School of Mathematics and Physics, University of Queensland, Australia}

\author{Kavan Modi}
\email{kavan@quantumlah.org}
\affiliation{School of Physics and Astronomy, Monash University, Australia}

\author{Thomas M. Stace}
\affiliation{ARC Centre of Excellence in Engineered Quantum Systems and\\ School of Mathematics and Physics, University of Queensland, Australia}

\begin{document}
\maketitle

\begin{abstract}
Quantum error correction (QEC) is essential for fault-tolerant quantum computation. Often in QEC errors are assumed to be independent and identically distributed and can be discretised to a random Pauli error during the execution of a quantum circuit. In real devices, however, the noise profile is much more complex and contains non-trivial spatiotemporal correlations, such as cross-talk, non-Markovianity, and their mixtures. Here, we examine the performance of two paradigmatic QEC codes in the presence of complex noise by using process tensors to represent spatiotemporal correlations beyond iid errors. This integration is an instance of the recently proposed \textit{strategic code}, which combines QEC with process tensors. In particular, we construct the maximum likelihood (ML) decoder for a quantum error correction code with a process tensor. To understand the computational overhead and implications of this approach, we implement our framework numerically for small code instances and evaluate its performance. 
We also propose a method to evaluate the performance of strategic codes and construct the ML decoder with an efficient tensor network approximation. Our results highlight the possible detrimental effects of correlated noise and potential pathways for designing decoders that account for such effects.

\end{abstract}

\newpage
\section{Introduction}

Achieving fault tolerance is a crucial stepping stone towards building a quantum computer. To this end, there has been a recent push to experimentally realise quantum error correction (QEC), with highly sophisticated implementations~\cite{Siddiqi, Sivak, Wallraff, google_2023,google_2024}, recently reviewed in~\cite{campbell}. Quantum error correcting codes are designed to protect the encoded quantum information against noise due to interaction between the quantum computer and its environment, and necessarily make simplifying assumptions about the statistics of the noise. Namely, the evaluation of QEC codes is commonly based on the strong assumption that the error is independent and identically distributed (iid) and can be discretized to a random Pauli error during the execution of a quantum circuit. However, real devices are subject to cross-talk and non-Markovianity, which violate the iid assumption~\cite{PhysRevX.9.021045,White-NM-2020, Sarovar2020, white2021many, Parrado2021, White2022-jq, PRXQuantum.3.020335, white2023filtering, PRXQuantum.4.040311, White2023-eq}.

At the same time, researchers have developed new theoretical and experimental techniques to characterise and control more realistic quantum noise, i.e., noise with correlations that lead to errors beyond the iid model. In particular, the process tensor framework~\cite{Pollock2018, Pollock2018a, Milz2020, Milz2021-pj} has emerged as a comprehensive representation of spatiotemporal noise correlations on a device. Recently, Ref.~\cite{Tanggara2024-an} combined the process tensor framework with quantum error correction, dubbed the \textit{strategic code}, to enable the study of QEC in the presence of more general noise. This framework covers general cases of QEC, including Floquet codes, and finds optimal QEC codes under the constraint of a given process tensor using a semi-definite program.

In this paper, we work within the same framework, to construct the maximum likelihood (ML) decoder for a QEC code under non-iid noise profiles represented by a process tensor. Then, we evaluate the performance of our framework by numerical experiment with the five-qubit perfect code~\cite{Laflamme1996-ur} and the seven-qubit Steane code~\cite{Steane}. Our paper highlights the connections between natural representations of complex quantum noise, quantum error correcting codes, and QEC decoders. 

The process tensor framework parameterises noise affecting the system accurately, but this comes with a higher computational cost. In this work, we also propose a scalable method to evaluate the performance of strategic codes and construct optimal decoders by employing tensor network approximation methods, particularly Matrix Product States (MPS)~\cite{perez-garcia_Matrix_2007a, schollwoeck_densitymatrix_2011}. We represent the process tensor and tester in an efficient MPS form by automatically detecting the bias in the outcomes from syndrome measurements. Applying our method to the Steane code, we numerically confirm that, in low-noise regimes, our approach can construct a decoder with accuracy comparable to that of exact contraction results in a much shorter time.

\section{Quantum Stochastic Processes and Complex Quantum Noise}

We start by introducing the tools required to describe multi-time quantum correlations. This includes channels, measurements, instruments, the Choi representation, and process tensors.

\subsection{Vectorisation of operator matrices}
First,  we introduce  \textit{vectorisation} of operator matrices~\cite{Gilchrist2009-vl, Watrous2018-td}.
Vectorisation is a mapping from a space of linear operators $\Linearmap{\Hilbertspace_{\mathrm{o}}}{\Hilbertspace_{\mathrm{i}}}$ to $\Hilbertspace_{\mathrm{o}}\otimes\Hilbertspace_{\mathrm{i}}$ for any choice of Hilbert spaces.
Let the standard basis of $\Linearmap{\Hilbertspace_{\mathrm{o}}}{\Hilbertspace_{\mathrm{i}}}$ be denoted by 
\begin{equation}\label{eq:elem_basis}
\{\standerdmatrixbasis{i,j}=\ketbra{i}{j}\},
\end{equation}
that maps a vector of computational basis $\ket{j} \in \Hilbertspace_{\mathrm{i}}$ to a vector of computational basis $\ket{i} \in \Hilbertspace_{\mathrm{o}}$.
Vectorisation is the linear map
\begin{align}
    \vectorisation: \Linearmap{\Hilbertspace_{\mathrm{o}}}{\Hilbertspace_{\mathrm{i}}}\rightarrow \Hilbertspace_{\mathrm{o}}\otimes\Hilbertspace_{\mathrm{i}}\qquad
\text{with action}
\qquad
\vectorisation(\standerdmatrixbasis{i,j}) = \ket{i}\otimes \ket{j} =: \doubleket{\standerdmatrixbasis{i,j}}.
\end{align}
Here, the $\doubleket{\cdot}$ notation emphasises that the vector represents a matrix and is often called double-ket. 

For example, the density matrix can be represented as a vector
\begin{align}
\rho = \sum_{rs} \rho_{rs} \standerdmatrixbasis{r,s} \rightarrow \doubleket{\rho}:=\sum_{rs} \rho_{rs} \doubleket{\hat{\Delta}_{r,s}},
\end{align}
In general, it holds by linearity that 
\begin{align}
    \vectorisation(\ketbra{\psi}{\phi}) = \ket{\psi}\otimes\ket{\phi^*},
\end{align}
for $\ket{\psi}\in\Hilbertspace_{\mathrm{o}}$ and $\ket{\phi}\in \Hilbertspace_{\mathrm{i}}$, where $*$ denotes the complex conjugate. The double bra $\doublebra{\cdot}$ also can be defined by taking the adjoint $\dagger$, generalising the Bra-Ket notation,
\begin{align}
    \doublebra{\standerdmatrixbasis{i,j}} := \left(\doubleket{\standerdmatrixbasis{i,j}}\right)^{\dagger}.
\end{align}
Vectorisation is just the vectorised representation of matrices. 
The inner product of the vectorised matrices is the Hilbert-Schmidt inner product on the operator space:
\begin{align}
\ev{\vectorisation(A),\vectorisation(B)} = \tr(A^\dagger B) =: \langle\!\langle A|B\rangle\!\rangle
\end{align}
for all $A, B\in \Linearmap{\Hilbertspace_{\mathrm{o}}}{\Hilbertspace_{\mathrm{i}}}$. Finally, the vectorisation mapping is fully reversible, i.e., we can map the vectorised operator back to its matrix form.

\subsection{Quantum Channels}

Vectorisation is particularly useful to express a quantum channel, $\check{\mathcal{E}}$, that maps a density matrix $\doubleket{\rho}$ to another density matrix $\doubleket{\rho'}$ as matrix operating on a vector:
\begin{align}
    \check{\mathcal{E}}\doubleket{\rho} 
    =\sum_{r's',rs} \check{\mathcal{E}}_{r's',rs}\rho_{rs} \doubleket{\standerdmatrixbasis{r',s'}} =: \doubleket{\rho'}.
\end{align}
The matrix form of the map $\mathcal{E}$ is then
\begin{align}
    \check{\mathcal{E}} :=\sum_{r',s',r,s} \check{\mathcal{E}}_{r',s',r,s} \doubleket{\standerdmatrixbasis{r',s'}}\doublebra{\standerdmatrixbasis{r,s}}.
\end{align}

\begin{figure}[htbp]
    \subfloat[\label{fig:CJI}]{\includegraphics[width=0.35\linewidth]{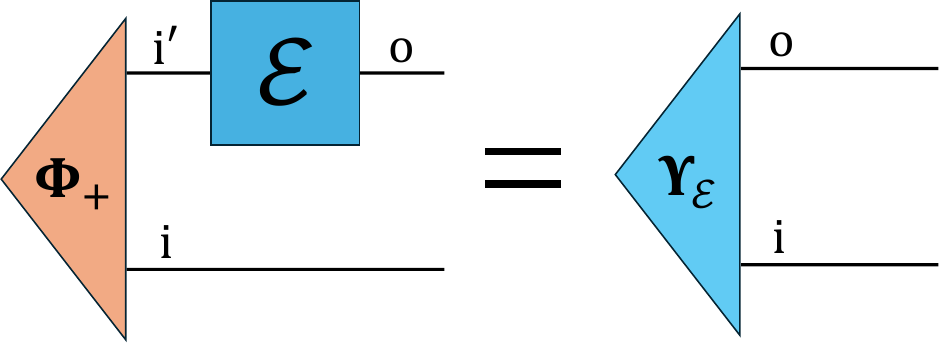}}
    \hspace{50pt}
    \subfloat[\label{fig:CJI_evolution}]{\includegraphics[width=0.35\linewidth]{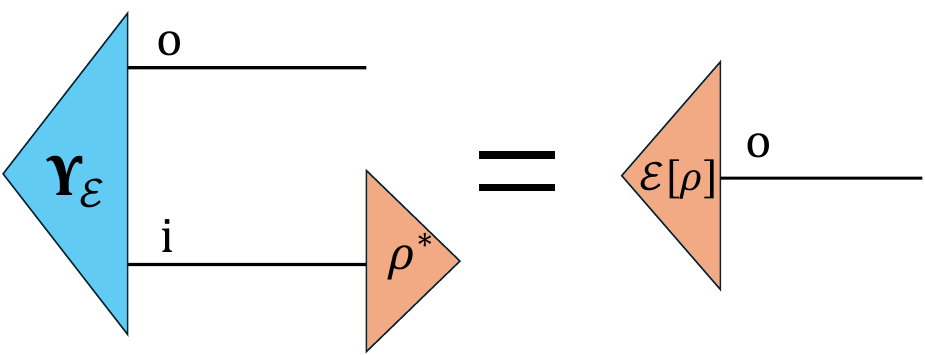}} \vspace{20pt} \\
    \subfloat[\label{fig:CPmap}]{\includegraphics[width=0.35\linewidth]{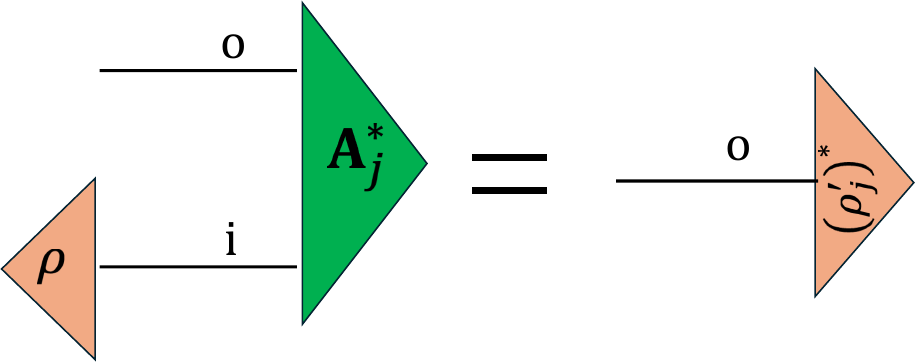}}
    \hspace{50pt}
    \subfloat[\label{fig:trace}]{\includegraphics[width=0.35\linewidth]{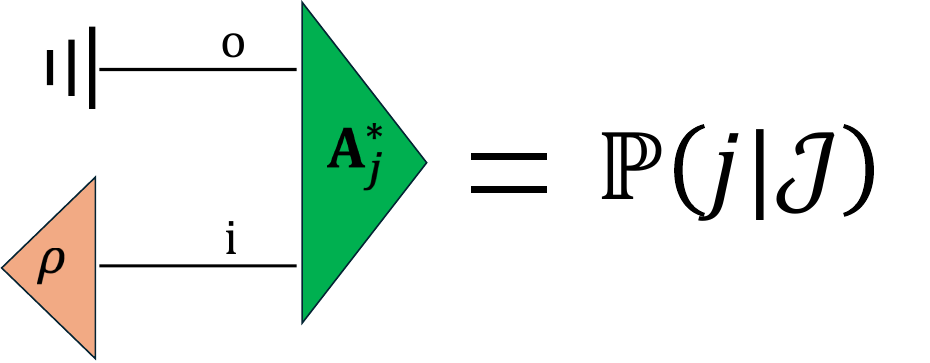}}
    \caption{(a)~The graphical representation of Choi-Jamiolkowski isomorphism. An orange triangle with $\Phi_{+}$ and a blue block with $\mathcal{E}$ on the left side of the equation are an unnormalised MES and a quantum Channel~$\mathcal{E}$. An blue triangle with $\Choimatrix{\mathcal{E}}$ is a Choi state of $\Choimatrix{\mathcal{E}}$. (b)~The input/output relation of Choi states. The orange triangle with $\rho^{*}$ is a vectorized density matrix~$\doublebra{\rho^{*}}$. Contraction of this tensor network becomes the vectorised density matrix after passing through $\mathcal{E}$. (c)~The action of a CP map $\mathcal{A}_j$. A green triangle with $\mathbf{A}^{*}_j$ denotes the Choi state of $\mathcal{A}_j$. If $\doubleket{\rho}$ is inputted, $\mathcal{A}_j$ will return the density matrix corresponding to the label $j$, i.e. $\rho'_j$. In this graphical representation, $\rho'_j$ returns to double-bra. Therefore, $(\rho'_j)^*$ is written in the triangle instead of $\rho'_j$. (d)~The three black lines on $\mathrm{o}$ denote the trace of $\Hilbertspace_\mathrm{o}$. In this case, it will be the probability that an instrument $\mathcal{J}$ returns an outcome $j$.}
\end{figure}

Another common representation of quantum channels is the so-called Choi form. The Choi-Jamiolkowski isomorphism enables us to treat a completely positive (CP) map as a Choi state, that is a positive semi-definite matrix.  This involves an isomorphism between $\mathscr{B}(\Hilbertspace_\textrm{i}) \to \mathscr{B}(\Hilbertspace_\textrm{o})$ and $\mathscr{B}(\Hilbertspace_{\textrm{i}}) \otimes \mathscr{B}(\Hilbertspace_\textrm{o})$, where $\mathscr{B}(\Hilbertspace)$ denotes the operator space of bounded linear operators on $\Hilbertspace$.
The Choi state $\Choimatrix{\mathcal{E}}$ corresponding to a quantum channel~$\mathcal{E}$ is constructed through the action of $\mathcal{E}$ on one half of an unnormalised maximally entangled state (MES)~$\ket{\Phi^+}=\sum_{i=0}^{d}\ket{ii}$ and the identity map $\mathcal{I}$ on the other half such as Fig.~\ref{fig:CJI}:
\begin{align}
\Choimatrix{\mathcal{E}} 
&= (\mathcal{E}\otimes\mathcal{I}) \left[\ketbra{\Phi^+}{\Phi^+}\right]
= \sum_{i,j=1}^{d} \mathcal{E}\left[\ketbra{i}{j}\right]\otimes \ketbra{i}{j},\label{eq:doubleket_channel}
\end{align}
where $\{\ket{i}\}$ is orthogonal basis of each Hilbert space, $\Hilbertspace_\textrm{i}$. The Choi matrix $\Choimatrix{\mathcal{E}} \in \mathscr{B}(\Hilbertspace_{\textrm{i}}) \otimes \mathscr{B}(\Hilbertspace_\textrm{o})$ is isomorphic to the quantum map $\mathcal{E}$. We depict the Choi matrix in Fig.~\ref{fig:CJI_evolution} and its action is written as 
\begin{align}
    \check{\mathcal{E}}\doubleket{\rho} &= \vectorisation(\tr_{\textrm{i}}\left[(\1_{\mathrm{o}}\otimes\rho^T)\Choimatrix{\mathcal{E}} \right]).
    = \ _\textrm{i}\langle\!\langle\rho^{*}|\Choimatrix{\mathcal{E}}\rangle\!\rangle_\textrm{io},
\end{align}
where $\tr_{\textrm{i}}[\cdot]$ is the trace over the input space $\Hilbertspace_i$, and $\1_{\mathrm{o}}$ denotes the identity matrix on output state $\Hilbertspace_{\mathrm{o}}$.

\subsection{Quantum measurement and instrument}

Quantum channels encompass generalised quantum measurements, i.e., positive operator valued measure (POVM) and quantum instruments. A POVM is defined as a collection of positive matrices $\mathcal{J} = \{E_j\}^n_{j=1}$ with the complete relation $\sum_j E_j =\1$. When a state $\rho$ is measured, POVM yields the outcome of the measurement and the probability of observing this outcome:
\begin{equation}
\probability{j|\mathcal{J}} = \tr(\rho E_j),
\end{equation}

The POVM does not tell us the post-measurement state of the system. A quantum instrument is a generalisation of POVM that enables us to compute the state transition corresponding to each measurement outcome, as well as the probability of obtaining each outcome. An instrument $\mathcal{J}$ is a collection of CP maps $\mathcal{J}=\{\mathcal{A}_j\}$ that add up a CPTP map~$\mathcal{A}=\sum_{j=1}^{n}\mathcal{A}_j$. 
The post-measurement state and the corresponding probability are given as
\begin{align}
\rho'_j = \mathcal{A}_j[\rho] = \sum_{\alpha}K_{\alpha,j} \ \rho \ K^{\dagger}_{\alpha,j} \qquad \mbox{and} \qquad
&    \probability{j|\mathcal{J}} = \tr(\mathcal{A}_j[\rho]).
\end{align}
The output state $\rho'$ is not normalized, while the trace of the input state $\rho$ is normalized. Above, the sum runs over the Kraus operators $K_{\alpha_j}$ that represent the CP map~$\mathcal{A}_j$. POVM operators are obtained from quantum instrument as $E_j = \sum_{\alpha} K^{\dagger}_{\alpha,j} K_{\alpha,j}$.

These two formalisms of a density matrix and the probability above are also representable in Choi states.
The action of a map $\mathcal{A}_j$ on a state $\rho$ is 
\begin{align}
    \rho'_{j} = \mathcal{A}_j[\rho] = \tr_{\mathrm{i}}\left[\mathbf{A}^{T}_j(\rho\otimes \1_{\mathrm{o}})\right] \quad \text{with probability} \quad
    \probability{j|\mathcal{J}} = \tr_{\mathrm{i,o}}\left[\mathbf{A}^{T}_j(\rho\otimes \1_{\mathrm{o}})\right],
\end{align}
where $\mathbf{A}_j\in \mathcal{B}(\Hilbertspace_{\mathrm{i}}\otimes \Hilbertspace_{\mathrm{o}})$ is the Choi state of $\mathcal{A}_j$ and $\rho \in \mathcal{B}(\Hilbertspace)$. In the double-ket description, as shown in Fig.~\ref{fig:CPmap} and \ref{fig:trace}, the above equation are
\begin{align}
    \doublebra{(\rho'_j)^{*}} = \langle\!\langle\mathbf{A}^{*}_{j}|{\rho}\rangle\!\rangle_{\mathrm{i}} \quad \text{with probability} \quad
    \probability{j|\mathcal{J}} = \langle\!\langle\mathbf{A}^{*}_{j}|{\1}\rangle\!\rangle_{\mathrm{o}}\otimes|{\rho}\rangle\!\rangle_{\mathrm{i}},
\end{align}
where $\doubleket{\1}_{\mathrm{o}}$ denotes the vectorised trace operation on space $\Hilbertspace_{\mathrm{o}}$.

\subsection{Multi-time statistics in quantum process}

\begin{figure}[t]
    \centering
    \subfloat[\label{fig:processtensor} The system-environment interaction model with CP maps~$\CPmaps{\outcomes{\timesteps{k-1}}}$]{\includegraphics[width=.65\linewidth]{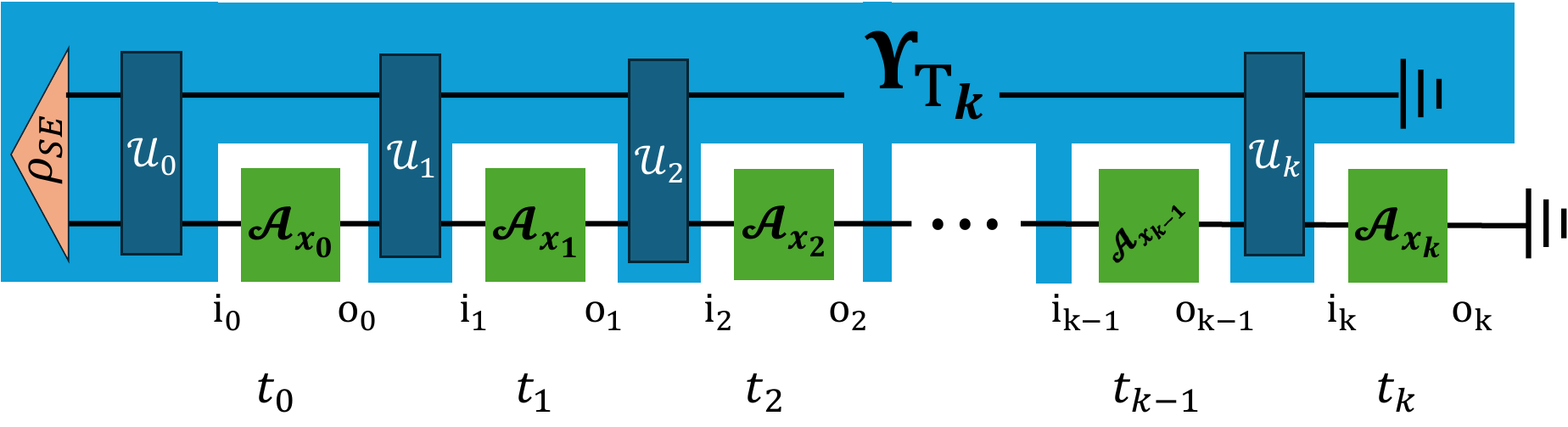}} \vspace{15pt} \\
    \subfloat[\label{fig:processtensor_init} The model if first CP map~$\mathcal{A}_{x_0}\in\CPmaps{\outcomes{\timesteps{k-1}}}$ is prepared to the state~$\rho$]{\includegraphics[width=0.55\linewidth]{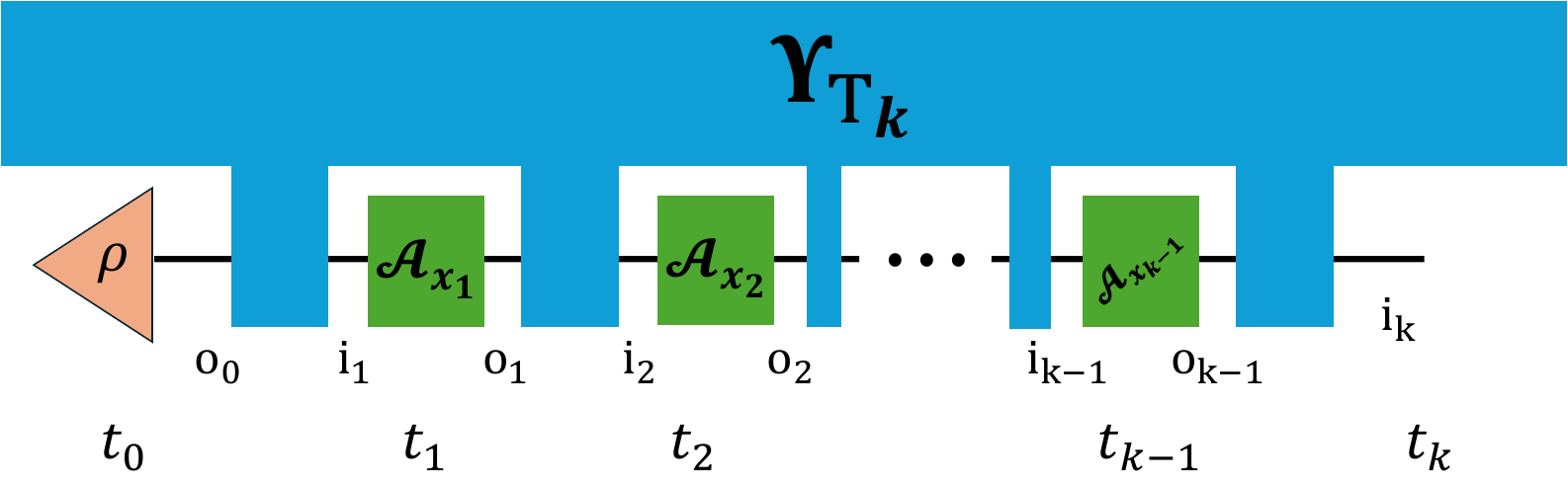}}\\
    \caption{\label{fig:sys_env_process}(a)~The system-environment interaction process. The orange triangle with $\rho_{SE}$ denotes the initial state. The darker blue boxes represent unitaries~$\mathcal{U}_j$. The green box represents the sequence of CP maps $\CPmaps{\outcomes{\timesteps{k-1}}}$. The initial state, unitaries and trace of $E$ can be seen a map~$\CPmaps{\outcomes{\timesteps{k-1}}}$ to the output state in $t_k$, that is $\rho[\CPmaps{\outcomes{\timesteps{k-1}}}] \in \mathcal{B}(\Hilbertspace_{\mathrm{i}_k})$. By tracing out the output state in $\mathrm{o}_k$, we will obtain the probability~$\probability{\textbf{x}_{{\timesteps{k}}}|\textbf{J}_{{\timesteps{k}}}}$. (b)~The process if the initial state is fixed in $t_0$. The process tensor enables us to treat not only the process that serves the fixed initial state, such as (a), but also the process that an experimenter can flexibly change the initial state. This process can be interpreted that the initial state $\rho_{SE}(t_0)$ is separable, i.e. $\rho_{SE}(t_0)=\rho_{E}(t_0)\otimes\rho$, or that the CP map $\mathcal{A}_{x_0}$ prepared the separable initial state $\rho$ in $t_0$. Moreover, the output of mapping can be seen as a quantum channel $\Choimatrix{\timesteps{k}}[\CPmaps{\outcomes{\timesteps{k-1}}}]$ from $\mathrm{o}_0$ to $\mathrm{i}_k$.}
\end{figure}

Now we generalise the above ideas to the multitime setting. We will work with the process tensor formalism~\cite{milz_Quantum_2021b}, which is used to describe correlated noise. To introduce the process tensor, let us define some sets below for convenience.
\begin{align}
&\timesteps{k} = \{t_0,t_1, \cdots, t_{k-1}, t_k\}, \quad
% \hfill(\textrm{temporal-steps})\\
\mathbf{J}_{\timesteps{k}} = \{\mathcal{J}_0, \mathcal{J}_1, \cdots, \mathcal{J}_k \},\\
% \hfill(\textrm{instruments on each temporal-step})\\
&\outcomes{\timesteps{k}} = \{ x_0, x_1, \cdots, x_k \}, \quad
% \hfill (\textrm{measurement outcomes on each temporal-step})\\
\CPmaps{\outcomes{\timesteps{k}}} = \{ \mathcal{A}_{x_0}, \mathcal{A}_{x_1}, \cdots, \mathcal{A}_{x_k} \},
% \hfill(\textrm{CP maps on each temporal-step})
\end{align}
where $\timesteps{k}$ is the set of times on which the process is defined and
$\mathbf{J}_{\timesteps{k}}$ is the the set of instruments on each temporal-step~$t_k$. These instruments are defined by outcomes $\outcomes{\timesteps{k}}$ and CP maps
$\CPmaps{\outcomes{\timesteps{k}}} $.

Consider the situation in which a $k$-step process is driven by a sequence of CP maps $\CPmaps{\outcomes{\timesteps{k-1}}} = \{\mathcal{A}_{x_0},\mathcal{A}_{x_1},\cdots\mathcal{A}_{x_{k-1}}\}$ of control operations after which we will obtain a quantum state $\rho(\CPmaps{\outcomes{\timesteps{k-1:0}}})$ conditioned on the choice of actions. This state will be represented by the interaction process with the environment $E$:
\begin{align}
    \rho[\CPmaps{\outcomes{\timesteps{k-1}}}] = \tr_E\left[\bigcirc^{k-1}_{j=0}\mathcal{U}_{j}\circ\mathcal{A}_{x_{j}}(\rho_{SE}(0))\right] =: \Choimatrix{\timesteps{k}}[\CPmaps{\outcomes{\timesteps{k-1}}}],
\end{align}
where $\rho_{SE}(0)$ denotes the initial state of the system~$S$ and environment~$E$, $\bigcirc$ denotes the composition of maps and $\mathcal{U}_{j}(\cdot) = U_j(\cdot)U^{\dagger}_j$ denotes the interaction with $E$ at $t_k$.
The CP maps act on just the $S$, but unitaries $\mathcal{U}_{j}$ act on both $S$ and $E$.
In this formalism, the CP maps can be selected independently from a sequence of instruments $\mathbf{J}_{\timesteps{k-1}}$. Thus, considering only the initial state, unitaries $\mathcal{U}_{j}$ and trace of $E$, it can be defined as a map from the control sequence~$\CPmaps{\outcomes{\timesteps{k-1}}}$ to the output state $\rho_k(\CPmaps{\outcomes{\timesteps{k-1}}})$. 
This is the process tensor $\Choimatrix{\timesteps{k}}$ such as shown in Fig.~\ref{fig:processtensor}.

\begin{figure}[tbp]
    \centering
    \subfloat[\label{fig:processtensor_dv}]{\includegraphics[width=0.16\linewidth]{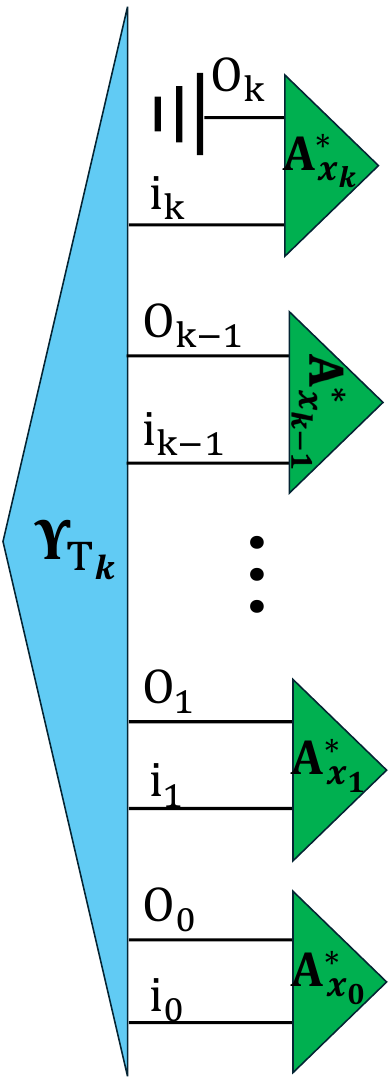}}
    \hspace{15pt}
    \subfloat[\label{fig:processtensor_init_dv} The vectorised representation of Fig.~\ref{fig:processtensor_init}.]{\includegraphics[width=0.5\linewidth]{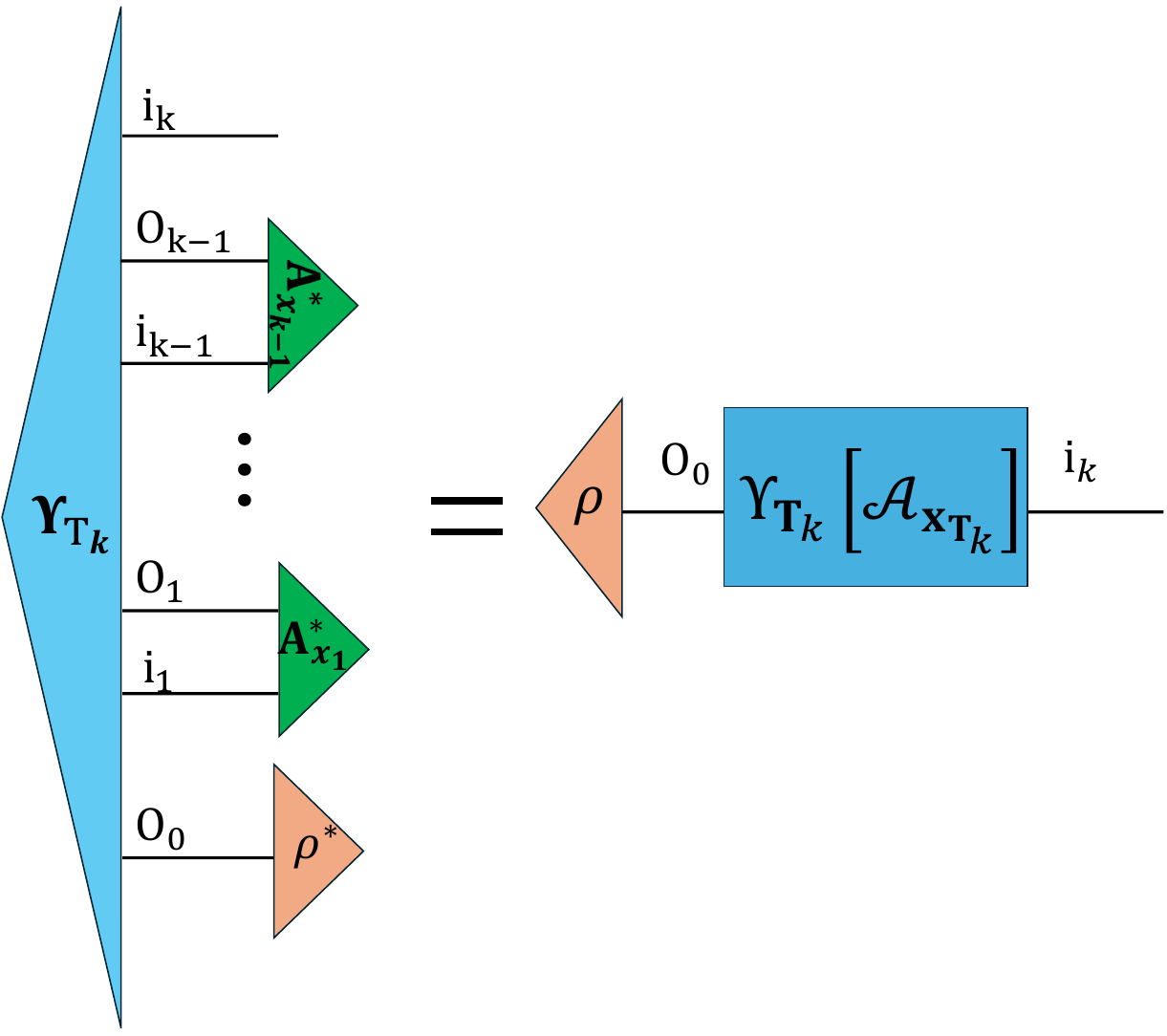}}
    \caption{(a)~The vectorised representation of Fig.~\ref{fig:processtensor}. The process tensor and CP maps can be written using vectorisation. (b)~The vectorised representation of Fig.~\ref{fig:processtensor_init}. The inner products of $\doubleket{\Choimatrix{\timesteps{k}}}$ with CP maps~$\CPmaps{\outcomes{\timesteps{k-1}}}$ is equal to a quantum channel of the initial state~$\rho$ to $\mathcal{B}(\Hilbertspace_{\mathrm{i}_k})$.}
\end{figure}

Suppose a given set of CP maps to be $\CPmaps{\outcomes{\timesteps{k}}}$ and the process tensor on $\timesteps{k}$ to be $\Upsilon_{\timesteps{k}}$, and consider that the instruments $\mathbf{J}_{\timesteps{k}}$ act on the process tensor $\Upsilon_{\timesteps{k}}$ and return the outcomes $\outcomes{\timesteps{k}}$, which means that the instruments affect the process as $\CPmaps{\outcomes{\timesteps{k}}}$.
The quantum state corresponding to the given results $\outcomes{\timesteps{k}}$ will be given by contraction between $\Upsilon_{\timesteps{k}}$ and $\CPmaps{\outcomes{\timesteps{k}}}$ such as
\begin{align}
    \Upsilon_{{\textbf{T}_k}}[\CPmaps{\outcomes{\timesteps{k}}}] = \mathrm{tr}_{\mathrm{i}_0,\mathrm{o}_0,\dots,\mathrm{i}_{k-1},\mathrm{o}_{k-1}, i_k}[\mathbf{A}_{\outcomes{\timesteps{k}}}^{T}\Upsilon_{\timesteps{k}}].
\end{align}
It can be written by vectorised representation as
\begin{align}
    \doubleket{\Upsilon_{\timesteps{k}}[\CPmaps{\outcomes{\timesteps{k}}}]} = (\doublebraket{\mathbf{A}_{\outcomes{\timesteps{k}}}^{*}}{\Choimatrix{\timesteps{k}}})^{T}.
\end{align}
This give us the probability corresponding to $\outcomes{\timesteps{k}}$ such as Fig.~\ref{fig:processtensor} and \ref{fig:processtensor_dv}:
\begin{align}
\probability{\textbf{x}_{{\timesteps{k}}}|\textbf{J}_{{\timesteps{k}}}} := \tr\left[\Choimatrix{\timesteps{k}}[\CPmaps{\outcomes{\timesteps{k}}}]\right] = \doublebraket{\1_{\mathrm{o}_k}}{\Upsilon_{\timesteps{k}}[\CPmaps{\outcomes{\timesteps{k}}}]},
\end{align}
where $\doubleket{\1}_{\mathrm{o}_k}$ denotes the vectorised identity operator of the output state on $\mathrm{o}_k$.
The density matrix of step $t_k$ is also given as
\begin{align}
    \Upsilon_{{\timesteps{k}}}[\CPmaps{\outcomes{\timesteps{k-1}}}] &=: \rho(t_k|\outcomes{\timesteps{k-1}}, \mathbf{J}_{\timesteps{k-1}})\nonumber\\
&= \tr_{\mathrm{i}_0,\mathrm{o}_0,\dots,\mathrm{i}_{k-1},\mathrm{o}_{k-1}}[\mathbf{A}_{\outcomes{\timesteps{k-1}}}^{T}\Choimatrix{\timesteps{k}}],
\end{align}
where $\mathbf{A}_{\outcomes{\timesteps{k-1}}} = \mathbf{A}_{x_k}\otimes\cdots\otimes\mathbf{A}_{x_0}$ represents the Choi states corresponding to a sequence of independent CP maps $\CPmaps{\outcomes{\timesteps{k-1}}}$.
Its vectorisation is also given as
\begin{align}
    \doubleket{\rho(t_k|\outcomes{\timesteps{k-1}}, \mathbf{J}_{\timesteps{k-1}})}
= \doublebraket{\mathbf{A}_{\outcomes{\timesteps{k-1}}}}{\Upsilon_{\timesteps{k}}},
\end{align}
which is written graphically in Fig.~\ref{fig:processtensor_init_dv}.

\begin{figure}[tbp]
    \centering
    \subfloat[\label{fig:tester}The tester of $\timesteps{k}$]{\includegraphics[width=0.7\linewidth]{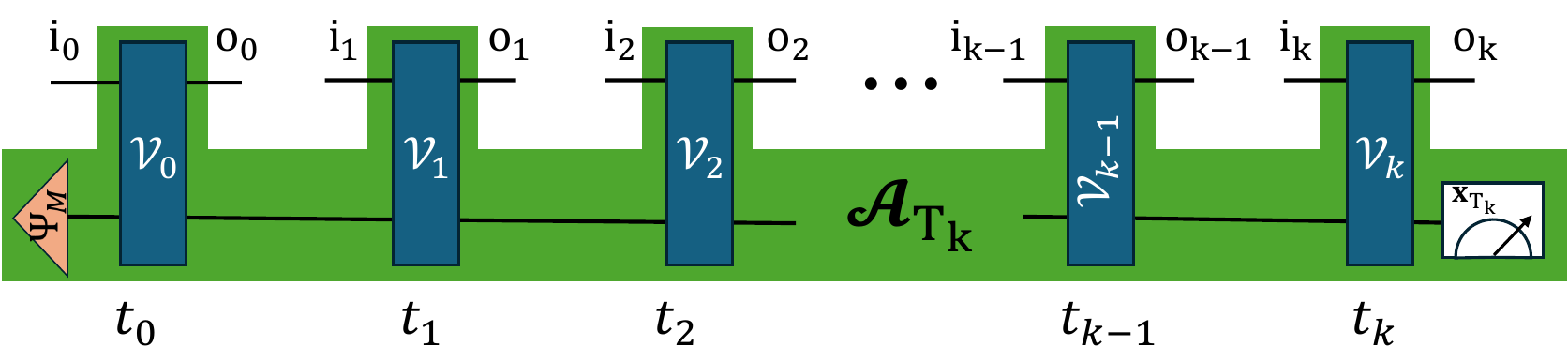}}\\
    \subfloat[\label{fig:processtensor_tester} The contraction of a process tensor~$\Upsilon_{\timesteps{k}}$ and a tester~$\CPmaps{\timesteps{k}}$]{\includegraphics[width=0.7\linewidth]{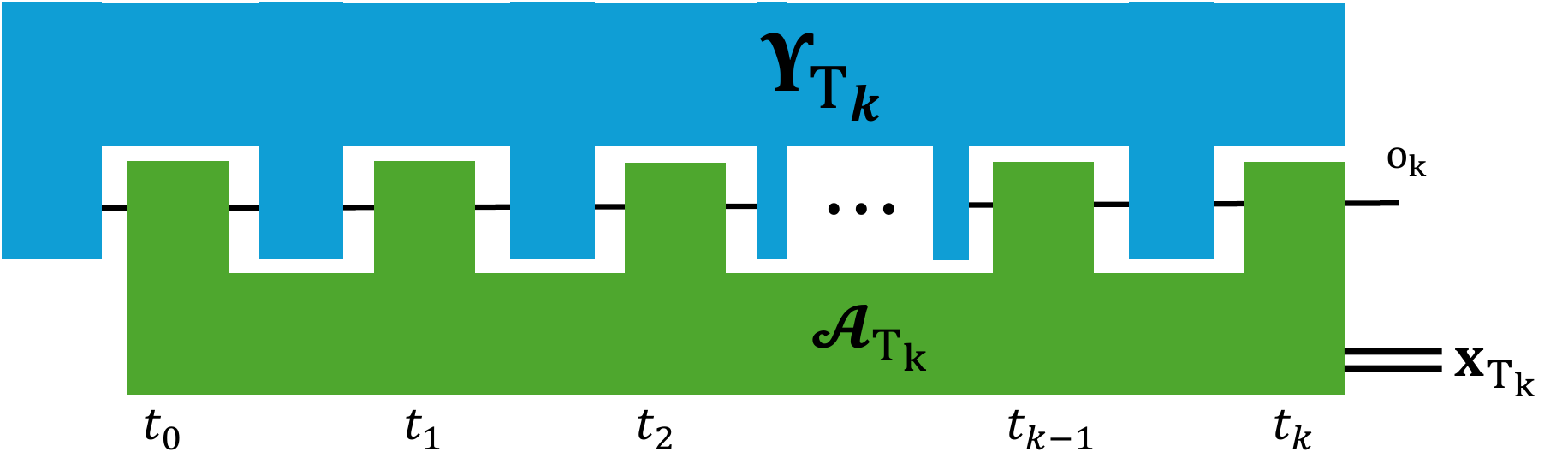}}\\
    \caption{(a)~The tester for time steps~$\timesteps{k}$. The upper-black line represents the quantum channel for the system, and the under-black line represents the ancillary system of the memory. The orange triangle represents the initial state of the memory system~$\Psi_M$, which will be finally measured and return the outcomes~$\mathbf{x}_{\timesteps{k}}$. The blue boxes represent unitaries through which the memory interacts with the system. (b)~The process tensor can be seen as the map of the tester as well as $\CPmaps{\outcomes{\timesteps{k}}}$. We can obtain the outcomes~$\outcomes{\timesteps{k}}$ as classical information, which is denoted as a double line of the left-hand side of a tester~$\CPmaps{\timesteps{k}}$ in the figure.}
\end{figure}

The discussion so far showed the case of the independent instruments for the process tensor.
The instruments~$\mathbf{J}_{\timesteps{k-1}}$ involving some correlations are also permitted, named \textit{tester}~\cite{Milz2021-pj}.
The experimenter would be able to condition the choice of an instrument at time $t'_j$ on the outcomes they recorded at times $t_j<t'_j$ using a tester. This conditioning would lead to a classical temporal correlation on the ``instrument", which is practised, for example, in the syndrome measurements and the decoding of QEC.
The tester tensor can be represented as the instrument having ``common memory," which technically is represented by ancillary qubits in Fig.~\ref{fig:tester}.

The Choi state of a tester is always expressed for such correlated operations using a local linear basis~$\{\hat{\mathbf{A}}_{x_j}\}$ at $t_{j}$ as 
\begin{align}
    \mathbf{A}_{\outcomes{\timesteps{k}}} = \sum_{\outcomes{\timesteps{k}}} \alpha_{\outcomes{\timesteps{k}}} \bigotimes_{j=0}^{k} \hat{\mathbf{A}}_{x_j},
\end{align}
where $\alpha_{\outcomes{\timesteps{k}}}$ are generic coefficient that can be non-positive.

The process tensor framework can be interpreted as the process tensor representing noise on an actual device, and the sequence of instruments represents the control sequence for the quantum device.
For example, the experimenter can apply some quantum gates to qubits, but the qubits might also interact with the environment, which no one can notice during the experiment.
The process tensor enables us to describe these hidden interactions. 
In addition, the process tensor can be estimated by process tensor tomography~\cite{White2022-jq, White2023-eq} from actual devices.
Thus, if the process tensor is given, we can simulate the control sequence as some instruments~$\mathbf{J}$.

\section{Quantum Error Correction in Presence of Complex Noise}

We now integrate the process tensor framework within quantum error correction to allow for complexity in noise, i.e. correlated noise. The real noise profile in quantum hardware has non-trivial spatiotemporal correlations~\cite{PhysRevX.9.021045,White-NM-2020, Sarovar2020, white2021many, Parrado2021, White2022-jq, PRXQuantum.3.020335, white2023filtering, PRXQuantum.4.040311, White2023-eq}. However, conventional QEC has generally targeted Markovian noise, such as iid noise.
Moreover, correlated noise can have detrimental effects on the performance of the QECC~\cite{PhysRevLett.97.040501, PhysRevLett.110.010502, Fowler2014-ue, Chubb2021-in, kam2024}. QEC considering the temporal correlation with a process tensor was recently named \textit{strategic code}~\cite{Tanggara2024-an}. Below we consider strategic code with the stabiliser formalism followed by considering decoding in this setting.

\subsection{Strategic code of the stabiliser codes}

Consider a pair of $[[n,k,d]]$ stabiliser code and its decoder $(\code,\decoder)$~\cite{Gottesman1997-bw, Nielsen2000-es}, where stabilisers and logical operators are elements of $\Pauligroup{n}$, the $n$-qubit Pauli group. 
The group~$\Pauligroup{n}$ contains all operators of the form $z\bigotimes^{n}_{j=1}\sigma^{i_j}$, with complex coefficient $z\in \{\pm1, \pm i\}$. Here, we are denoting the Pauli operators, $\sigma^{0}=\1$, $\sigma^{1}=X$, $\sigma^{2}=Y$ and $\sigma^{3}=Z$.
The codespace of $\code$ is determined by the stabilisers, an abelian group of Pauli operators~$\mathcal{S}_{\code}\subset\Pauligroup{n}$. Every state $\ket{\psi}$ in the codespace satisfies $S\ket{\psi}=\ket{\psi}$ for all stabilizers~$S\in\mathcal{S}_{\code}$.
$\code$ is specified by the stabilizer generators
\begin{align}
    \stabilizers_{\code} = \{G_1, G_2, \cdots, G_{n-k}|^\forall i\in\{1,2,\cdots,n-k\}\},
\end{align}
where $G_i\in \mathcal{S}_{\code}$ is $n-k$ independent elements of stabilizer $\mathcal{S}_{\code}$.
% \com{What's the formula saying?  Are $G_i$ the generators?}
The logical operators of $\code$ generate a non-abelian group $\logicals_{\code}\subset\Pauligroup{n}$. 
A set of generators for the group consists of the $k$ $Z$-type and $k$ $X$-type logical operators.
We denote these as $X_a$ and $Z_a$ with $a \in \{1,\cdots,k\}$.
These operators commute with all stabilizers in $\mathcal{S}_{\code}$, while anti-commuting pairwise, i.e., $X_a Z_b = (-1)^{\delta_{a,b}}Z_b X_a$.

Here, the syndrome measurements are a set of projective measurements corresponding to each element in $\stabilizers_{\code}$,
\begin{align}
    \mathbf{J}_{\code} = \left\{\mathcal{J}_j = \left\{\projector_{x_j} = \frac{\1 +(-1)^{x_j} G_j}{2}\right\} \right\}^{{n-k}}_{j=0},
\end{align}
where $x_j\in\{0,1\}$. We have expressed these in the language of instruments. The CP map of syndrome measurement at $t_j$ is represented as 
\begin{align}\label{eq:syndrome_measurements}
    \Pi_{x_j}(\bullet) = \projector_{x_j} \bullet \projector_{x_j},
\end{align}
and let us say its Choi state is $\choiprojector_{x_j}$ and the measurement sequence corresponding to the outcomes~$\outcomes{\timesteps{1:n-k}} \in\{0,1\}^{n-k}$ is $\choiprojector_{\outcomes{\timesteps{1:n-k}}} = \bigotimes_{j=1}^{n-k} \choiprojector_{x_j}$.

Suppose an encoding state of $\code$ is $\rho_{L,{\mathbf{i}}}$, then 
the output state through the measurement sequence~$\Pi_{\outcomes{\timesteps{1:{n-k}}}}$ with a process tensor $\Choimatrix{\timesteps{n-k+1}}$ would be written as
\begin{flalign}
    \doubleket{\rho(t_{n-k+1}|\outcomes{\timesteps{1:n-k}}, \mathbf{J}_{\code})}=\left(\doublebra{\choiprojector_{\outcomes{\timesteps{1:n-k}}}}\otimes\doublebra{\rho^{*}_{L,{\mathbf{i}}}}\right)\doubleket{\Choimatrix{\timesteps{n-k+1}}}.
\end{flalign}
After acquiring the syndrome outcome~$\outcomes{\timesteps{1:n-k}}$, the decoder~$\decoder$ will give us a recovery operation that is denoted as
\begin{align}\label{eq:recovery}
    \recovery(\rho|\outcomes{\timesteps{1:n-k}}) = R(\outcomes{\timesteps{1:n-k}}) \rho R^{\dagger}(\outcomes{\timesteps{1:n-k}})
\end{align}
and it enables us to fix errors in the data qubits.
Suppose the Choi state of the recovery operation is $\mathbf{R}({\outcomes{\timesteps{1:n-k}}})$, the output logical state will be
\begin{flalign}
    \doubleket{\rho_{L,{\textrm{o}}}} =
    \left(\doublebraket{\mathbf{R}^{*}({\outcomes{\timesteps{1:n-k}}})}{\rho (t_{n-k+1}|\outcomes{\timesteps{1:n-k}}, \mathbf{J}_{\timesteps{1:{n-k}}})}\right)^{T}.
\end{flalign}
The syndrome measurements and the recovery process will be graphically represented as in Fig.~\ref{fig:strategiccode}.
The sequence of syndrome measurements~$\Pi_{\outcomes{\timesteps{1:n-k}}}$ and the recovery operation~$\recovery(\cdot|\outcomes{\timesteps{1:n-k}})$ can be seen as the larger tester because the classically temporal correlation exists as syndrome outcomes~$\outcomes{\timesteps{1:n-k}}$.
The Choi state of this larger tester will be written as
\begin{align}\label{eq:code_tester}
    \mathbf{C}_{\outcomes{\timesteps{1:n-k}}} &=\sum_{\outcomes{\timesteps{1:n-k}}'\in\{0,1\}^{n-k}} \choiprojector_{\outcomes{\timesteps{1:n-k}}}\otimes\mathbf{R}({\outcomes{\timesteps{1:n-k}}'})\delta_{\outcomes{\timesteps{1:n-k}}, \outcomes{\timesteps{1:n-k}}'},
\end{align}
where $\delta_{\outcomes{\timesteps{1:n-k}}, \outcomes{\timesteps{1:n-k}}'} = \prod_{(x_{t_j}, x'_{t_j})\in(\outcomes{\timesteps{1:n-k}}, \outcomes{\timesteps{1:n-k}}')} \delta_{x_{t_j}, x'_{t_j}}$, that is the Kronecker delta for measurement outcomes, which causes the correlation.

Now, we have integrated QEC with the process tensor. We will call this type of QEC scheme the \textit{interleave} QEC, which is more complicated than the phenomenological QEC but is not more primitive than the circuit simulation of QEC.
Next, we construct a decoder~$\decoder$ for the interleave QEC.
The following section will show the construction scheme of a maximum likelihood (ML) decoder~\cite{Ferris2014-du, Bravyi2014-jd, Chubb2021-zk} and a tensor-network simulation of the logical failure rate instead of a Monte Carlo simulation.

\subsection{Maximum likelihood~(ML) decoder}

Decoding errors are the most crucial issue on QEC.
The performance of QEC is decided by both the QEC code and its decoding.
The optimal decoder for QEC is the Maximum likelihood~(ML) decoder~\cite{Ferris2014-du, Bravyi2014-jd, Chubb2021-zk}, which finds the error correction operator that is most likely to return the system to the correct codespace depending on the syndromes~\cite{Iyer2015-lc, Farrelly2021-yx, Farrelly2022-xz}.
Here, we will show the construction method of the ML decoder from the process tensor and a stabiliser code~$\code$.

To construct the ML decoder, we will first define the \textit{pure error} (in other words, destabilisers).
The pure error is an abelian group of Pauli operators $\mathcal{P}_{\code}\subset\Pauligroup{n}$.
This group is defined by $n-k$ operators $P_i$ which satisfy anti-commuting with $G_j\in\stabilizers_{\code}$, i.e., $P_iG_j = (-1)^{\delta_{i,j}}S_jP_i$ and $P^2_i=\1$.
By measuring the error syndrome~$x_j$ corresponding to each $\mathcal{J}_j$, we obtain the syndrome outcomes~$\outcomes{\timesteps{1:n-k}}=\vec{s}\in\mathbf{s}$, which is a length-$(n-k)$ binary vector~$(s_1,s_2,\cdots,s_{n-k})^{T}$ as shown in the previous section.
 Here, $\mathbf{s}$ is the set of the possible binary vectors.
These outcomes indicate that $s_i=0$ if the error commuted with stabiliser~$G_i$ and $s_i=1$ if the error anti-commuted with $G_i$.
We can always find an error consistent with a given syndrome~$\vec{s}$ by using pure errors, i.e., the operator
\begin{align}
    P(\vec{s})= \prod_{i\in\{1,2,\cdots,n-k\}} P^{s_i}_{i}
\end{align}
give rise to syndrome~$\vec{s}$.
However, quantum codes are degenerate, which means some different errors can raise the same syndrome, for example, $P(\vec{s})$ and $P(\vec{s})S$ for any stabiliser $S\in \mathcal{S}_{\code}$ will give rise to the same syndrome~$\vec{s}$.
These degenerate errors with syndrome~$\vec{s}$ can be written as $P(\vec{s})SL$, where $S\in \mathcal{S}_{\code}$ and $L \in \mathcal{L}_{\code}$. The important thing is that all $P(\vec{s})SL$ has the same action on the codespace regardless of the choice of $S$.
On the other hand, the choice of different logicals~$L$ will have crucially different actions on the codespace.
ML decoders estimate the likelihood recovery operator~$L$ by optimizing a given objective function.
% To find the optimal correction, \add{ML decoders estimate the likelihood logical recovery operator~$L$ optimizing given objective function.
For example, the objective function for Pauli errors is given as the probability distribution
\begin{align}\label{eq:chi}
    \chi(L, \vec{s}) = \sum_{S\in\mathcal{S}_{\code}} \probability{P(\vec{s})SL}
\end{align}
for each logical $L\in\mathcal{L}_{\code}$, where $\probability{P(\vec{s})SL}$ is the probability that the error $P(\vec{s})SL$ occurred.
The best recovery operator is given by
\begin{align}
R(\vec{s}) = \bar{L}P(\vec{s}) \qquad \text{where}
\qquad \bar{L} = \argmax{L} \chi(L, \vec{s}).
\end{align}
Thus, the recovery operation will be given as 
\begin{align}
\recovery(\rho|\vec{s},\bar{L}) &= \left(\bar{L}P(\vec{s})\right) \rho \left(\bar{L}P(\vec{s})\right)^{\dagger}
= \bar{L}P(\vec{s}) \rho P(\vec{s})\bar{L}\label{eq:ML_recovery}\\
&= \mathcal{E}_{\bar{L}}\circ\mathcal{E}_{P(\vec{s})}(\rho),
\end{align}
where $\mathcal{E}_{P(\vec{s})}$ and $\mathcal{E}_{\bar{L}}$ denote the quantum channels of $P(\vec{s})$ and $\bar{L}$.
For uncorrelated error models, such as iid depolarising noise, the correction is obtained using tensor-network codes, where the probability of error occurrence factorizes for each data qubit~\cite{Farrelly2021-yx, Farrelly2022-xz}.

\begin{figure}
    \centering
    \subfloat{\includegraphics[width=.7\linewidth]{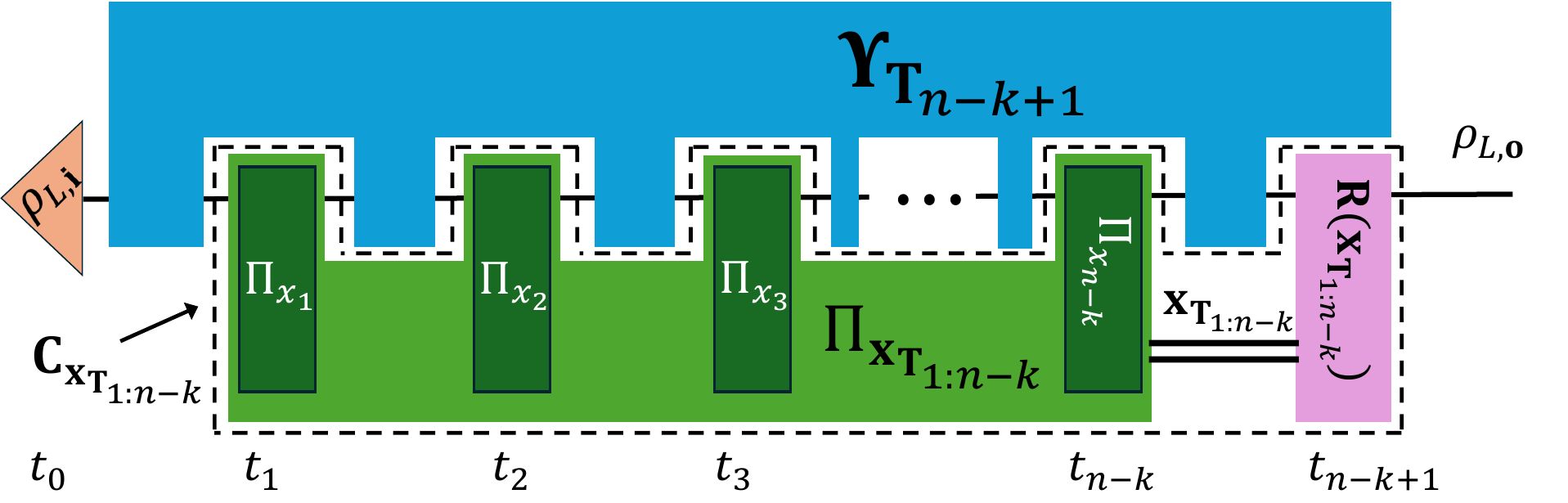}}
    \caption{The strategic code of a $[[n,k,d]]$ stabilizer code. The blue object represents the process tensor for $\timesteps{n-k+1}$ steps, the lighter green box represents the tester of syndrome measurements~$\choiprojector_{\outcomes{\timesteps{1:n-k}}}$, the darker green box represents each syndrome measurement~$\Pi_{x_j}$ at $t_j$, and the pink box represents the recovery operation~$\mathbf{R}(\outcomes{\timesteps{1:n-k}})$. By inputting the encoded state~$\rho_{L,\mathbf{i}}$ from the left-hand side, this tensor network outputs the logical output state~$\rho_{L,\mathbf{o}}$. The syndrome tester and the recovery operation are correlated classically through syndrome outcomes~$\outcomes{\timesteps{1:n-k}}$. Thus, the pair of them can be seen as the larger tester~$\mathbf{C}_{\outcomes{\timesteps{1:n-k}}}$.}\label{fig:strategiccode}
\end{figure}

To extend the maximum likelihood decoder to the interleaved noise processes, we represent the syndrome measurements and the recovery operation a tester tensor. We define the objective function through two different metrics of the ``closeness'' to the identity channel of the logical qubit. The first is the Hilbert-Schmidt inner product calculated by just taking the inner product of a given process tensor and the code tester of Eq.~\eqref{eq:code_tester}. The second is the distance between the entire strategic-code process and the identity channel.

For the former, we define the Choi state of the recovery operation of the ML decoder. From Eq.~\eqref{eq:ML_recovery}, the Choi state of the recovery operation~$\mathbf{R}(L, \outcomes{\timesteps{1:n-k}})$ of the ML decoder can be decomposed to two Choi states:
\begin{equation}
    \mathbf{R}(L, \outcomes{\timesteps{1:n-k}}) =\mathbf{L}\star\mathbf{P}(\outcomes{\timesteps{1:n-k}}),
\end{equation}
where $\mathbf{P}(\outcomes{\timesteps{1:n-k}})$ and $\mathbf{L}$ are the Choi state of pure error corresponding to $\outcomes{\timesteps{1:n-k}}$ and the Choi state of a logical operator~$\mathcal{E}_{\bar{L}}$.
Here, we use the \textit{link product}~$\star$ defined in Ref.~\cite{chiribella2009theoretical, Milz2021-pj}.
The link product translates a concatenated map onto its Choi state, e.g., if $\mathbf{A}$ and $\mathbf{C}$ denote the Choi states of maps $\mathcal{A}$ and $\mathcal{C}$, respectively, $D=C\star A$ is the Choi state of the concatenated map~$\mathcal{C}$ and $\mathcal{A}$.
Thus, the tester of Eq.~\eqref{eq:ML_recovery} can be written with parameters $(L,\outcomes{\timesteps{1:n-k}})$ as
\begin{align}\label{eq:code_tester_ML}
    \mathbf{C}_{L,\outcomes{\timesteps{1:n-k}}} = \sum_{\outcomes{\timesteps{1:n-k}}'\in\{0,1\}^{n-k}}\choiprojector_{\outcomes{\timesteps{1:n-k}}}\otimes\mathbf{L}\star\mathbf{P}({\outcomes{\timesteps{1:n-k}}'})\delta_{\outcomes{\timesteps{1:n-k}}, \outcomes{\timesteps{1:n-k}}'}.
\end{align}

To obtain the entire QEC process for the stabiliser code on the strategic code, consider attaching the encoding channel on the left-hand side of the process tensor in Fig.~\ref{fig:strategiccode}.
Suppose $\mathscr{H}_{\textrm{logical}}$ and $\mathscr{H}_{\mathbf{o}}$ to be the Hilbert spaces of the logical qubit and corresponding data qubits.
The projector of the encoding map $\mathcal{E}_{\mathrm{enc}}: \mathscr{H}_{\textrm{logical}}\to \mathscr{H}_{\mathbf{o}}$ can be written as 
\begin{align}\label{eq:encode}
    \Pi_{\textrm{enc}} = \ket{\psi_{L,0}}\bra{0_{L}} + \ket{\psi_{L,1}}\bra{1_{L}},
\end{align}
where $\ket{i_{L}}\in\mathscr{H}_{\textrm{logical}}$ and $\ket{\psi_{L,i}}\in\mathscr{H}_{\mathbf{o}}$ for $i\in\{0,1\}$.
Then, the encoding channel can be denoted by
\begin{align}
    \mathcal{E}_{\textrm{enc}}(\bullet) = \Pi_{\textrm{enc}}\bullet \Pi^{\dagger}_{\textrm{enc}}.
\end{align}
Conversely, the state vector on $\mathscr{H}_{\mathbf{i}}$ can be mapped to $\mathscr{H}_{\textrm{logical}}$ using Eq.~\eqref{eq:encode} as
\begin{equation}
    \mathcal{E}_{\textrm{dec}}(\bullet) = \Pi_{\textrm{enc}}^\dagger \bullet \Pi_{\textrm{enc}}.
\end{equation}
Let the Choi state of the encoding and decoding channels denote $\choiprojector_{\textrm{enc}}$ and $\choiprojector_{\textrm{dec}}$, respectively.

By fixing the syndrome outcome~$\outcomes{\timesteps{1:n-k}}=\vec{s}$ and logical operator $L$, the object after taking the inner product of  $\choiprojector_{\textrm{enc}}$, $\mathbf{C}_{L,\vec{s}}$ and $\Choimatrix{\timesteps{n-k+1}}$ will be a Choi state:
\begin{align}\label{eq:qec_channel}
\doublebra{\choiprojector^*_{\mathrm{enc}}}\doublebraket{\mathbf{C}^{*}_{L,\vec{s}}}{\Choimatrix{\timesteps{n-k+1}}}    &=\vectorisation\left(\tr_{\mathbf{o}_0,\mathbf{i}_1,\mathbf{o}_1,\dots,\mathbf{i}_{n-k},\mathbf{o}_{n-k},\mathbf{i}_{n-k+1}}\left[\choiprojector^{T}_{\mathrm{enc}}\mathbf{C}^{T}_{L,\vec{s}}\Choimatrix{\timesteps{n-k+1}}\right]\right)^{\dagger}\nonumber\\
   % &=\tr_{\mathbf{i}_1,\mathbf{0}_1,\dots,\mathbf{i}_{n-k},\mathbf{0}_{n-k},\mathbf{i}_{n-k+1}}\left[\choiprojector^{T}_{\vec{s},L}\otimes\mathbf{P}^{T}(\vec{s})\otimes\mathbf{L}^{T}\Choimatrix{\timesteps{n-k+1}}\right]\nonumber\\
   &=: \doublebra{\Choimatrix{\timesteps{n-k+1}}[\mathbf{C}_{L,\vec{s}}]\star \choiprojector_{\mathrm{enc}}}.
\end{align}

Suppose we input the encoded state to this channel. In that case, the output state may not return to the codespace depending on the error distribution. This is because we are considering temporal syndrome measurements that will sometimes fail to detect the occurrence of an error.
Consider the case where the error $e$, which is detectable by the instrument $\Pi_k$, occurs after step $k$. 
In this case, the error will not be detected and the recovery operation will not correct it.
We need to evaluate the closeness of the identity channel of code space and the strategic code process of $\code$ to calculate the logical failure rate of the code.
To achieve this, we apply perfect (noiseless) syndrome measurements followed by a decoder to the code space achieves:
\begin{align}\label{eq:qec_channel2}
    &\doublebra{\choiprojector^{*}_{\mathrm{enc}}}\doublebra{\mathbf{C}^{*}_{L,\vec{s}}}\doublebra{\mathbf{C}^{*}_{L_{\text{pdc}},\vec{s}_{\text{pdc}}}}\doublebraket{\choiprojector^{*}_{\mathrm{dec}}}{\Choimatrix{\timesteps{n-k+1}}}\doubleket{\mathbf{I}_{\timesteps{n-k+1:2(n-k+1)+1}}}\nonumber\\
    &=\vectorisation(\tr_{\mathbf{o}_0,\mathbf{i}_1,\dots,\mathbf{o}_{2*(n-k)},\mathbf{i}_{2*(n-k+1)}}[\choiprojector^{T}_{\mathrm{enc}}\mathbf{C}^{T}_{L,\vec{s}}\mathbf{C}^{T}_{L_{\text{pdc}},\vec{s}_{\text{pdc}}}\choiprojector^{T}_{\mathrm{dec}}\:
    \Choimatrix{\timesteps{n-k+1}}\mathbf{I}_{\timesteps{n-k+1:2(n-k+1)+1}}])^{\dagger}\nonumber\\
   &=: \doublebra{\choiprojector_{\mathrm{dec}} \star\mathbf{I}_{\timesteps{n-k+1}:2(n-k+1)+1}[\mathbf{C}_{L_{\text{pdc}},\vec{s}_{\text{pdc}}}]\star 
\Choimatrix{\timesteps{n-k+1}}[\mathbf{C}_{L,\vec{s}}] \star \choiprojector_{\mathrm{enc}}},
\end{align}
where the $\mathbf{I}_{\timesteps{n-k+1}:\timesteps{2(n-k+1)+1}}$ is the process tensor that bridges input and output by identity channels.

\begin{figure}
    \centering
    \includegraphics[width=0.5\linewidth]{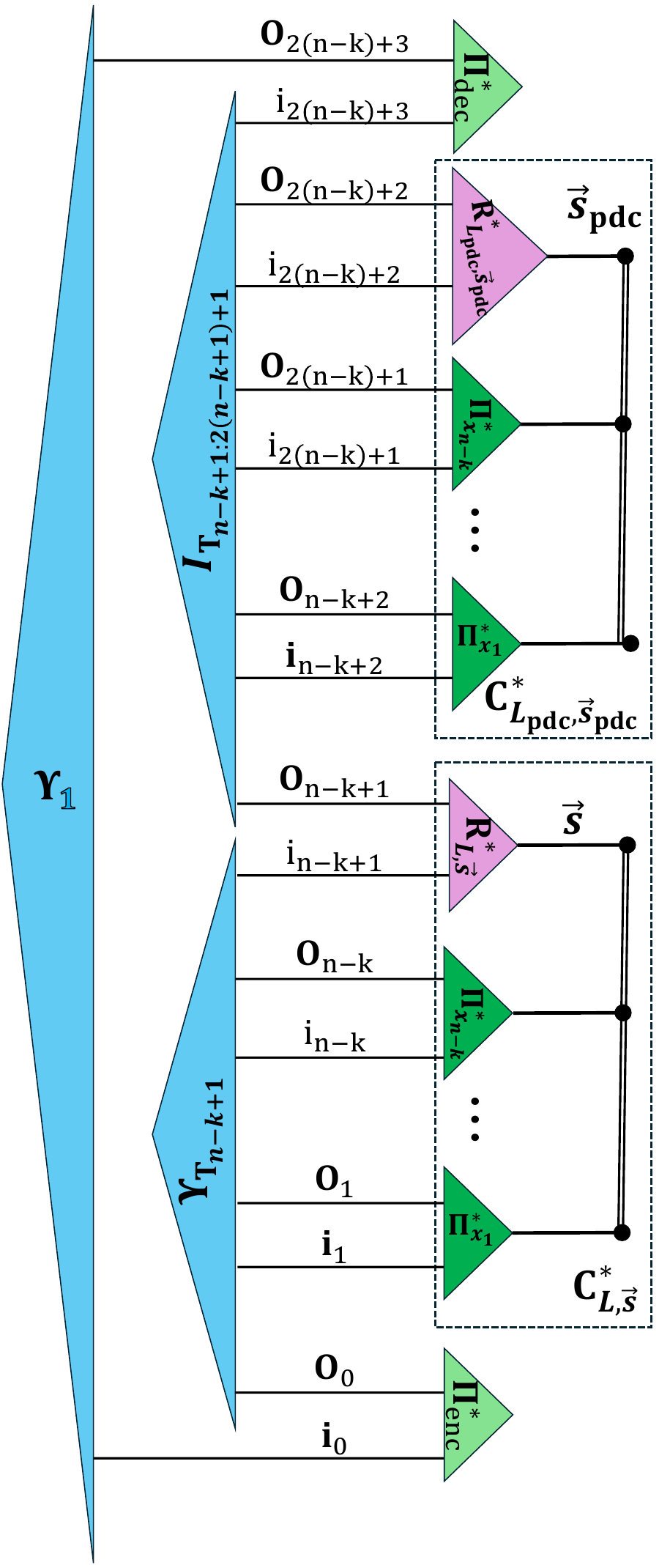}
    \caption{The graphical representation of the calculation of $\chi_{\textrm{HS}}(L, \vec{s})$ in Eq.~\eqref{eq:chi_contraction}. The green triangles represent the Choi state of projectors, the left blue triangle represents the Choi state of identity channel~$\1$ that is a $2^n\times 2^n$ matrix, the right blue triangles represent the Choi state of the process tensor~$\Choimatrix{\timesteps{n-k+1}}$ and process tensor of identity channel $\mathbf{I}_{\timesteps{n-k+1}:2(n-k)+2}$, and the pink triangle represents the decoder that depends on parameters of $L$ and $\vec{s}$. The wires from the green triangles feed forward the syndrome outcomes to the recovery operation on the pink triangle. The graphical representation clarifies the relationship of legs taking inner products in Choi states. This graphical representation can also seen as the tensor-network, that means $\chi_{\textrm{HS}}(L,\vec{s})$ is given by the contractions of legs on the TN of Choi states.}
    \label{fig:chi_TN}
\end{figure}

% From Eq.~\eqref{eq:doubleket_channel}, the Choi state of identity channel is 
% \begin{equation}
%     \Choimatrix{\1} = \sum_{i,j} \ketbra{i}{j}\otimes \ketbra{i}{j}.\label{eq:Choi_iden}
% \end{equation}
% and its vectorisation is denoted as $\doubleket{\Choimatrix{\1}}$.
%Considering the inner product of Eq.~\eqref{eq:qec_channel2} and Eq.~\eqref{eq:Choi_iden} on $\mathscr{H}_{\textrm{logical}}$, we 
We can measure the closeness of the identity channel and the strategic code process of~$\code$:
\begin{align}\label{eq:chi_contraction}
    \chi_{\textrm{HS}}(L, \vec{s}) &= 
    \doublebraket{\choiprojector_{\mathrm{dec}} \star\mathbf{I}_{\timesteps{n-k+1}:2(n-k+1)+1}[\mathbf{C}_{L_{\text{pdc}},\vec{s}_{\text{pdc}}}]\star 
\Choimatrix{\timesteps{n-k+1}}[\mathbf{C}_{L,\vec{s}}] \star \choiprojector_{\mathrm{enc}}}{\Choimatrix{\1}}
\end{align}
that can be interpreted graphically in Fig.~\ref{fig:chi_TN}. Above $\doubleket{\Choimatrix{\1}}$ is vectorised Choi state of identity channel on $\mathscr{H}_{\textrm{logical}}$. 
Here, we define $\chi_{\textrm{HS}} (L, \vec{s})$ as the success rate of the logical recovery operation for the first case.
Eq.~\eqref{eq:chi_contraction} can be used as the objective function for the interleaved noises with the process tensor.
The optimal recovery operation $\bar{L}(\vec{s})$ from $\chi_{\textrm{HS}}(L, \vec{s})$ gives us the operation with the highest likelihood of returning the qubits to the correct codespace
\begin{equation}
    \bar{L}(\vec{s}) = \argmax{L} \chi_{\textrm{HS}}(L, \vec{s}).
\end{equation}
As a consequence, the maximum value for each syndrome outcome $\vec{s}$ reflects the success probability of recovering to the codespace.
Thus, the logical failure rate of strategic code of~$\code$ is given by 
\begin{equation}
    p_{\textrm{fail}} = 1- \sum_{\vec{s}\in\mathbf{s}}\chi_{\textrm{HS}}(\bar{L}(\vec{s}), \vec{s}). \label{eq:ler_inner}
\end{equation}

The metric in Eq.~\eqref{eq:chi_contraction} does not depend on the coherent (or the off-diagonal) terms of the \textit{logical} quantum channel. We can get around this by employing another metric of the closeness, the distance between the identity channel for a logical qubit and the strategic code process Eq.~\eqref{eq:qec_channel2}.
Let us denote the probability of obtaining the series of syndrome outcomes $\outcomes{\timesteps{1:n-k}}=\vec{s}$ as
\begin{equation}
     p(\vec{s}) = \tr\left[\rho_L\choiprojector^{T}_{\mathrm{enc}}\mathbf{\Pi}^{T}_{\vec{s}}\Choimatrix{\timesteps{n-k+1}}\right]
\end{equation}
where $\rho_L$ represents the completely mixed state on $\mathscr{H}_{\textrm{logical}}$.
The distance between the identity channel for a logical qubit and the strategic code process Eq.~\eqref{eq:qec_channel2} is defined as
\begin{align}\label{eq:chi_contraction_distance}
\chi_{\textrm{CD}}(L, \vec{s}) = \left\|\choiprojector_{\mathrm{dec}} \star\mathbf{I}_{\timesteps{n-k+1}:2(n-k+1)+1}[\mathbf{C}_{L_{\text{pdc}},\vec{s}_{\text{pdc}}}]\star 
\Choimatrix{\timesteps{n-k+1}}[\mathbf{C}_{L,\vec{s}}] \star \choiprojector_{\mathrm{enc}} - \Upsilon_{\1}\right\|_{2},
\end{align}
where $\| \cdot \|_{2}$ and $\Upsilon_{\1}$ are the 2-norm and the Choi representation of identity operator on $\mathscr{H}_{\textrm{logical}}$, respectively~\cite{darmawan_TensorNetwork_2017}.
The recovery operation $\bar{L}(\vec{s})$ is given by
\begin{equation}
    \bar{L}(\vec{s}) = \argmin{L} \chi_{\textrm{CD}}(L, \vec{s}).
\end{equation}
% However, by using it as a weight for the probability of each syndrome outcome, the logical failure probability can be obtained without much difference from the evaluation with the fidelity $\chi_{\textrm{HS}}$.
The logical failure rate for the objective function $\chi_{\textrm{CD}}(\bar{L}(\vec{s}), \vec{s})$ is defined as
\begin{equation}
    p_{\textrm{fail}} =\sum_{\vec{s}\in\mathbf{s}} p(\vec{s}) \  \chi_{\textrm{CD}}(\bar{L}(\vec{s}),\vec{s}). \label{eq:ler_distance}
\end{equation}
However, it cannot satisfy the strict axioms of probability~\cite{darmawan_TensorNetwork_2017}.
Note that there are other possible metrics for evaluating the decoding error or the logical failure probability, such as the diamond distance~\cite{Nakata2022-qr} or other $p$-norm distances~\cite{Watrous2018-td}.

\section{Implementation}

We now consider two numerical implementations of the stabiliser strategic code. Our goal is to demonstrate the structural compatibility of stabiliser-based QEC and its decoder under complex noise as represented by the process tensor. In this Section, we first construct a noise model where we can tune the local noise parameters, cross-talk, and non-Markovianity. We then study the performance of the five-qubit code and the Steane code under this model. 
We also show that the tensor network approximation methods can be applied to the estimation of the code performance within the strategic code framework.

\begin{figure}[tbp]
    \centering
    \includegraphics[width=0.5\linewidth]{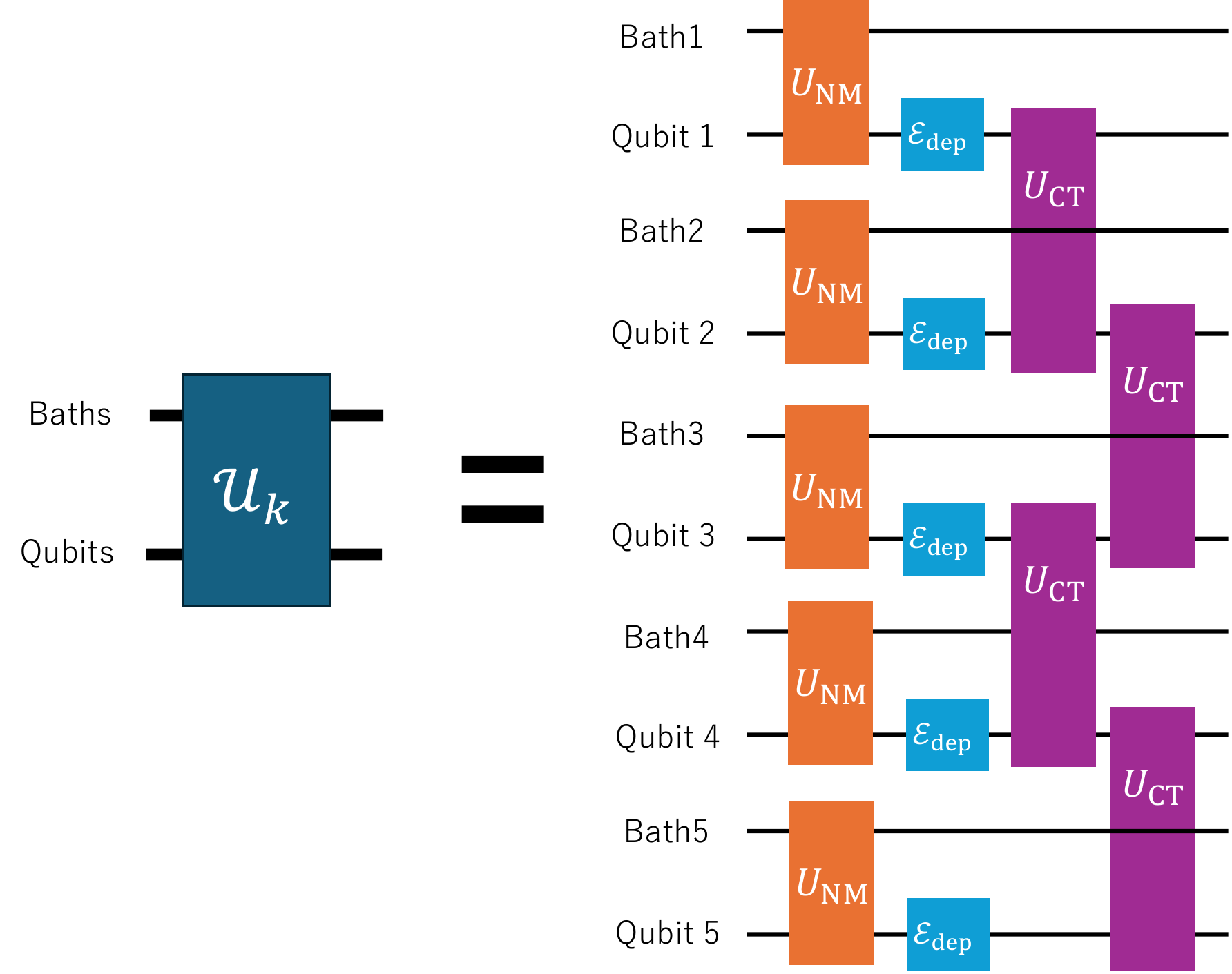}
    \caption{The interaction block $\mathcal{U}_k$ of our process tensor at $k$-th step. The blue blocks are the stochastic depolarising noise $\mathcal{E}_{\textrm{dep}}$, the orange blocks are the Heisenberg interaction $U_{\textrm{NM}}(J)$ to make non-Markovian noise, and the 
    purple blocks are the $ZZ$ crosstalk interaction $U_{\textrm{CT}}(J)$.
    \label{fig:process_tensor}}
\end{figure}

\subsection{Noise Model with Complex Features}

To evaluate the performance of our scheme and the effect of spatiotemporal non-Markovian noise, we performed numerical experiments of our process with the five-qubit code with the following noise model. Here, we allow for three types of errors:
\begin{enumerate}
    \item Stochastic noise: a single qubit depolarising error:
    \begin{equation}\label{eq:depolarising}
    \mathcal{E}_{\textrm{dep}}(\rho, p_{\textrm{err}}) = (1-p_{\textrm{err}}) \rho + p_{\textrm{err}}\sum_{\sigma \in \{X, Y, Z\}} \sigma\rho \sigma^{\dagger}.
    \end{equation}
    This is the standard iid Markovian noise.
    \item Heisenberg interaction process for non-Markovian noise:
    \begin{equation}\label{eq:NM_channel}
        U_{\textrm{NM}} \coloneqq \exp\left\{-i J_{\textrm{NM}} (X_iX_{B_i} + Y_iY_{B_i} + Z_iZ_{B_i})\right\}.
    \end{equation}
    Here, $i$ denotes the $i$-th data qubit of the system and $B_i$ is its local bath environment.
    
    \item Crosstalk interaction generated by a $ZZ$ coupling:
    \begin{equation}\label{eq:CT_channel}
        U_{\textrm{CT}} \coloneqq \exp\left\{-iJ_\mathrm{CT} \ Z_iZ_{i+1}\right\}.
    \end{equation}
    Here, both $i$ and $i+1$ denote data qubits on the system.
\end{enumerate}
    
\begin{figure}[tbp]
    \centering
    \subfloat[\label{fig:NM_processtensor} Non-Markovian process tensor for five qubit code]{
    \includegraphics[width=0.75\linewidth]{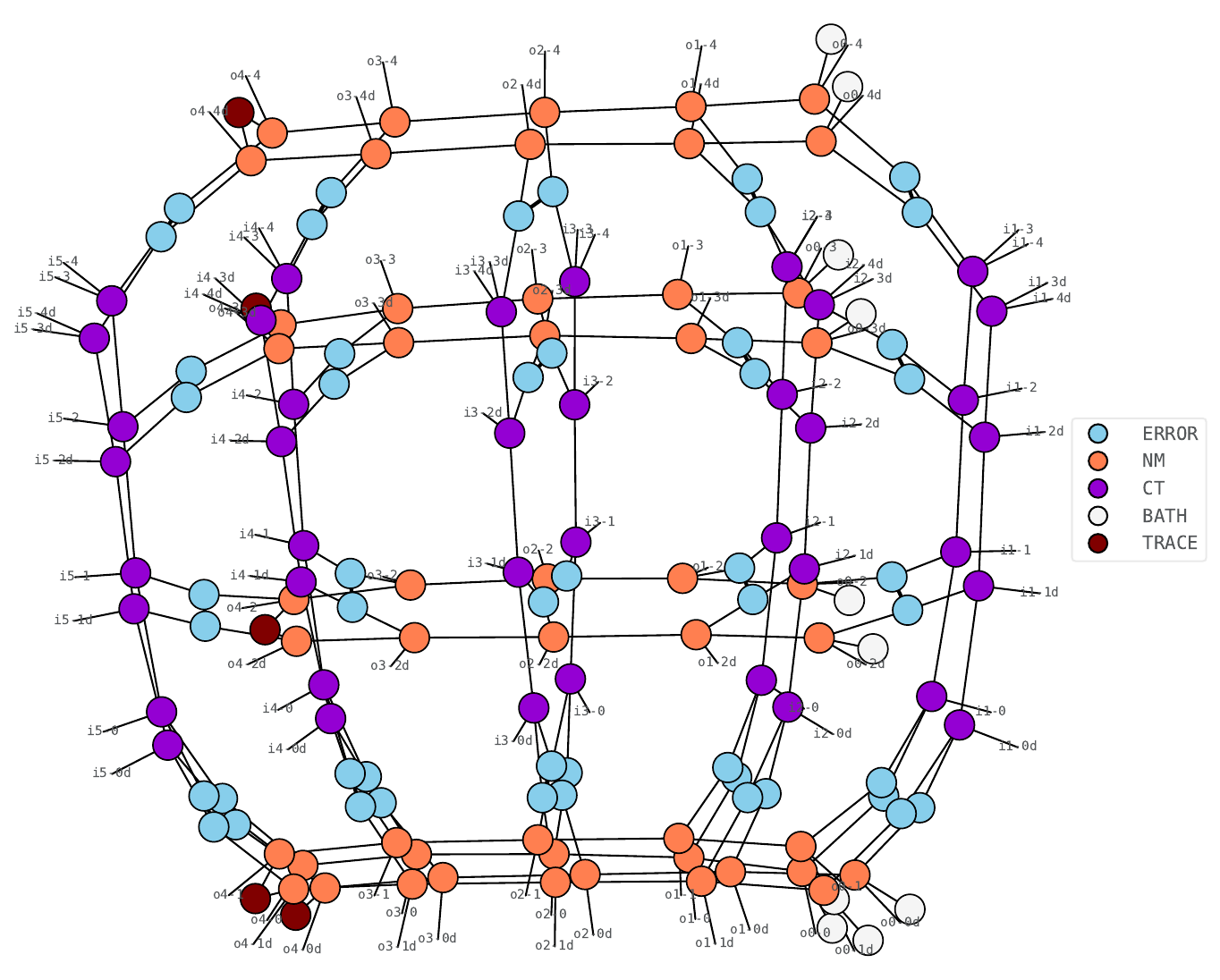}
    }\\
    \subfloat[\label{fig:Whole_process} Tensor-Network of whole process]{\includegraphics[width=0.85\linewidth]{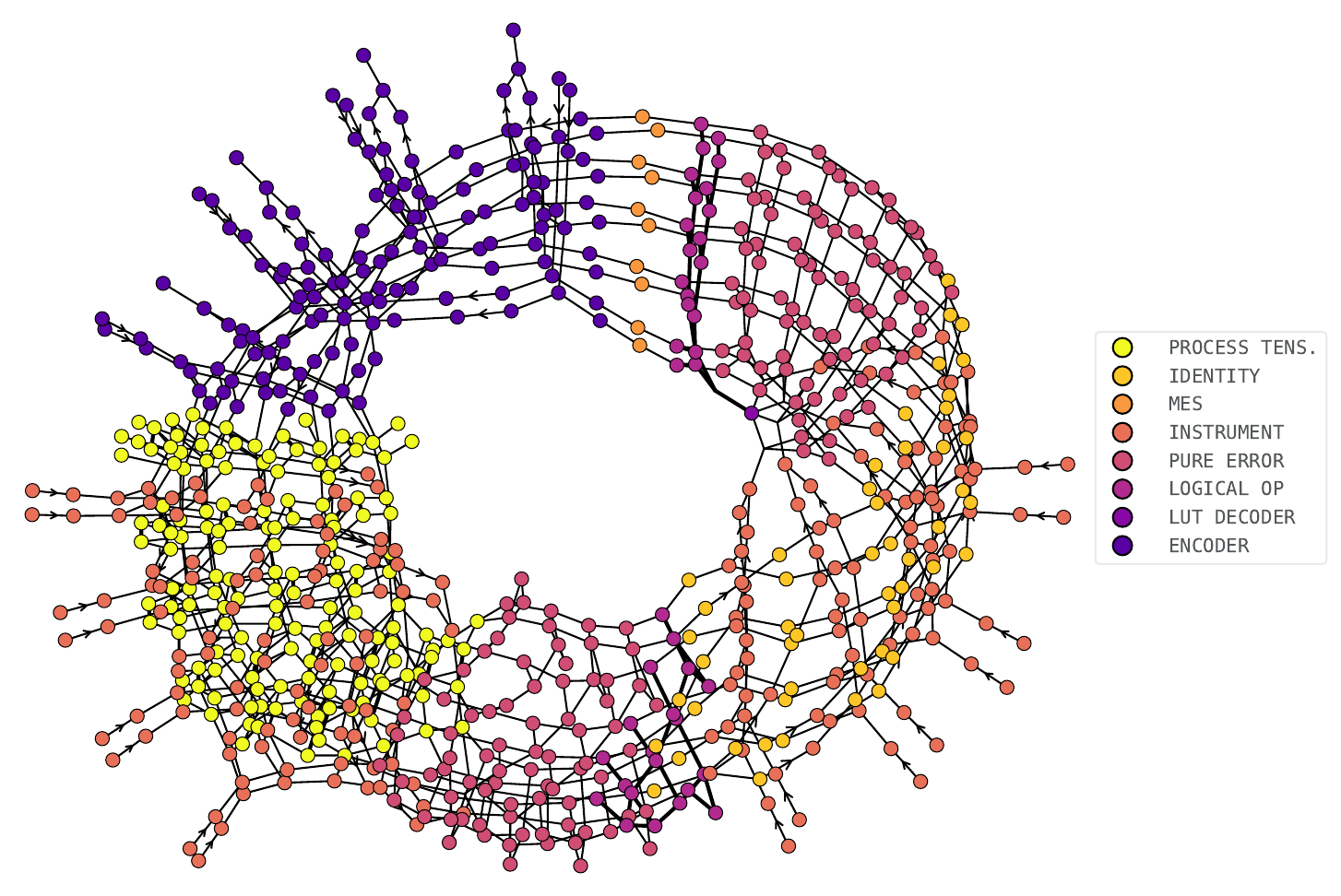}}
\caption{(a) The tensor-network representation of the process tensor.
ERROR, NM and CT on the legend are depolarising error channel Eq.~\eqref{eq:depolarising}, Non-Markovian interaction of Eq.~\eqref{eq:NM_channel}, and the cross-talk interaction Eq.~\eqref{eq:CT_channel}. BATH and TRACE are the baths on the process tensor and the trace of its mode. 
(b) The tensor-network representation of $\chi_{\textrm{HS}}(L, \vec{s})$. 
Each dot beside the LUT DECODER dot represents tensor objects with up to 3 legs.
The process tensor dots involve 
dots on (a). Identity dots represent the process tensor $\mathbf{I}_{\timesteps{n-k+1}:2(n-k+1)+1}$
are the instruments of syndrome measurements.
MES dot is a maximally entangled state representing the Choi state of the identity channel on data qubits.
Instrument dots represent four steps of syndrome measurements for each of process tensors $\Choimatrix{\timesteps{n-k+1}}$ and $\mathbf{I}_{\timesteps{n-k+1}:2(n-k+1)+1}$. Pure error and logical recovery operator dots are applied twice at each end of syndrome measurements.
The contraction of this tensor network returns the success probability of decoding as a matrix of dangling legs on the first four-round syndrome measurements and logical recovery operation.
LUT DECODER dot represents the five-qubit code look-up table decoder.}
\end{figure}

The $k$-th step dynamics, containing all of the above errors, is shown in Fig.~\ref{fig:process_tensor}. The corresponding process tensor is constructed by applying these interaction blocks sequentially and tracing out the environment qubits such as~Fig.~\ref{fig:sys_env_process}, and its tensor network representation is Fig.~\ref{fig:NM_processtensor}.

\subsection{Five-Qubit Code}
\label{sec:five-qubit}

The five qubit code is a $[[5,1,3]]$ stabiliser code with four stabilisers and two logical operators~\cite{Laflamme1996-ur}. The stabiliser generators, logical operators, and the pure errors of the five-qubit code are shown in Table.~\ref{tab:five_qubit_code}.

To obtain the logical failure rate along with Eq.~\eqref{eq:ler_inner} and \eqref{eq:ler_distance}, we represent the entire strategic-code process as a tensor network.
The entire tensor network of $\chi_{\textrm{HS}}$ is represented graphically as Fig.~\ref{fig:Whole_process} with the process flowing clockwise.
% The strategic-code process goes along with the clockwise direction in Fig.~\ref{fig:Whole_process}. 
The tensor legs along with this direction represent the Hilbert space of data qubits.
% \com{i don't understand the last sentence.}
The process starts at encoding channels (denoted as ENCODER), then the data qubits are exposed to noises as the process tensor, followed by syndrome measurements. The measured syndrome outcomes feed-forward to recovery operators consisting of PURE ERROR and LOGICAL OP. The process so far is connected to the instruments of syndrome measurements and the recovery operators again as $\mathbf{C}_{L_{\text{pdc}}, \vec{s}_{\text{pdc}}}$. Then, the input legs and output legs are stitched with the maximally entangled states. Note that we reduced $\choiprojector_{\mathrm{dec}}$ by merging it and $\choiprojector_{\mathrm{enc}}$ to ENCODER tensors.
The tensor network is implemented on quimb, a Python library for tensor network~\cite{gray2018quimb}.
We optimised the contraction order using HyperOptimizer on cotengra~\cite{gray_Hyperoptimized_2021}. The numerical calculation runs on a machine with an AMD EPYC 7532 32-core Processor and an NVIDIA A100 40GB GPU.

\begin{figure}
\centering
\subfloat[\label{fig:Non-Markovian_lfr} Effect of NM error]{%
  \includegraphics[width=0.57\linewidth]{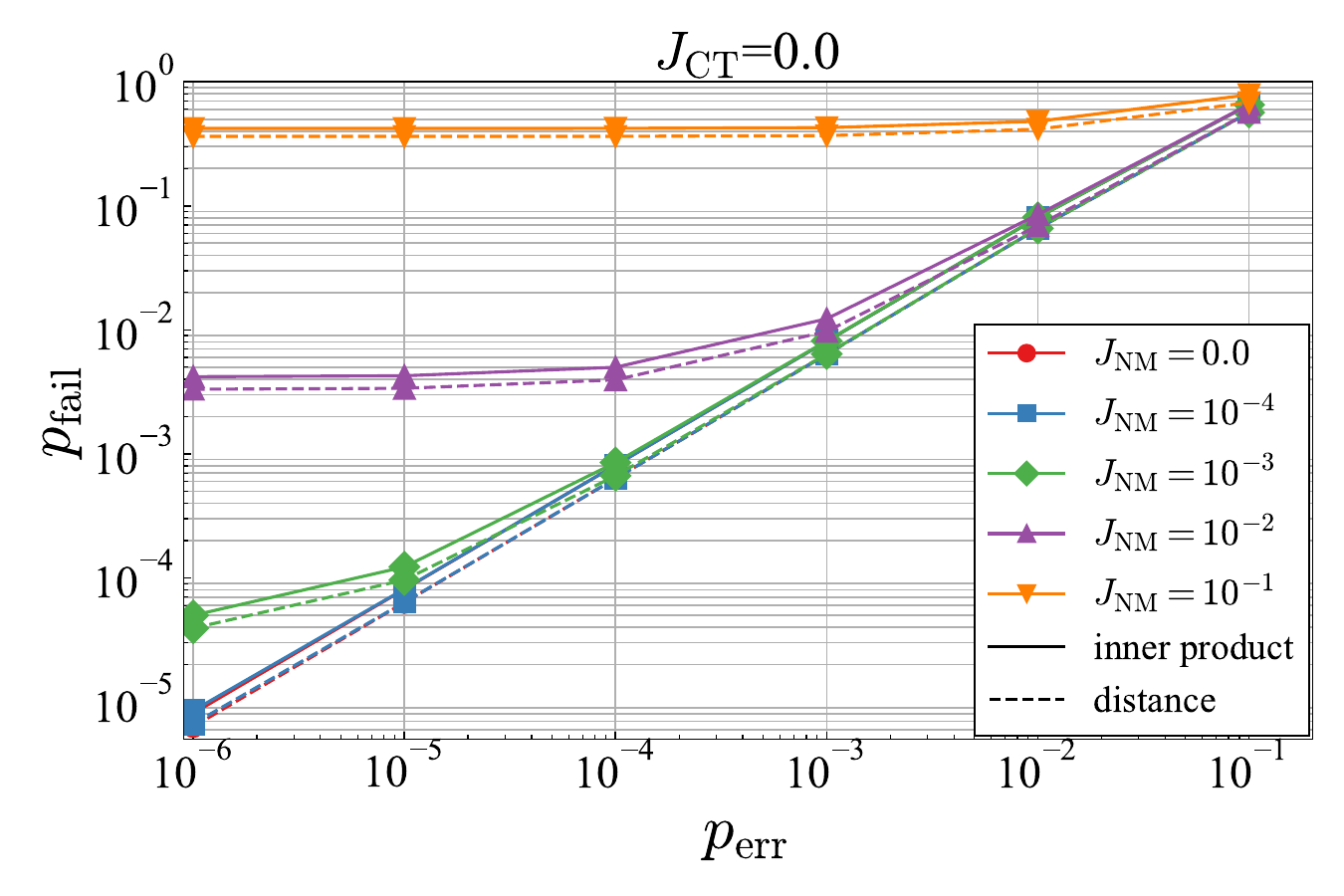}
}\\
\subfloat[\label{fig:cross_talk_lfr} Effect of CT error]{%
  \includegraphics[width=0.57\linewidth]{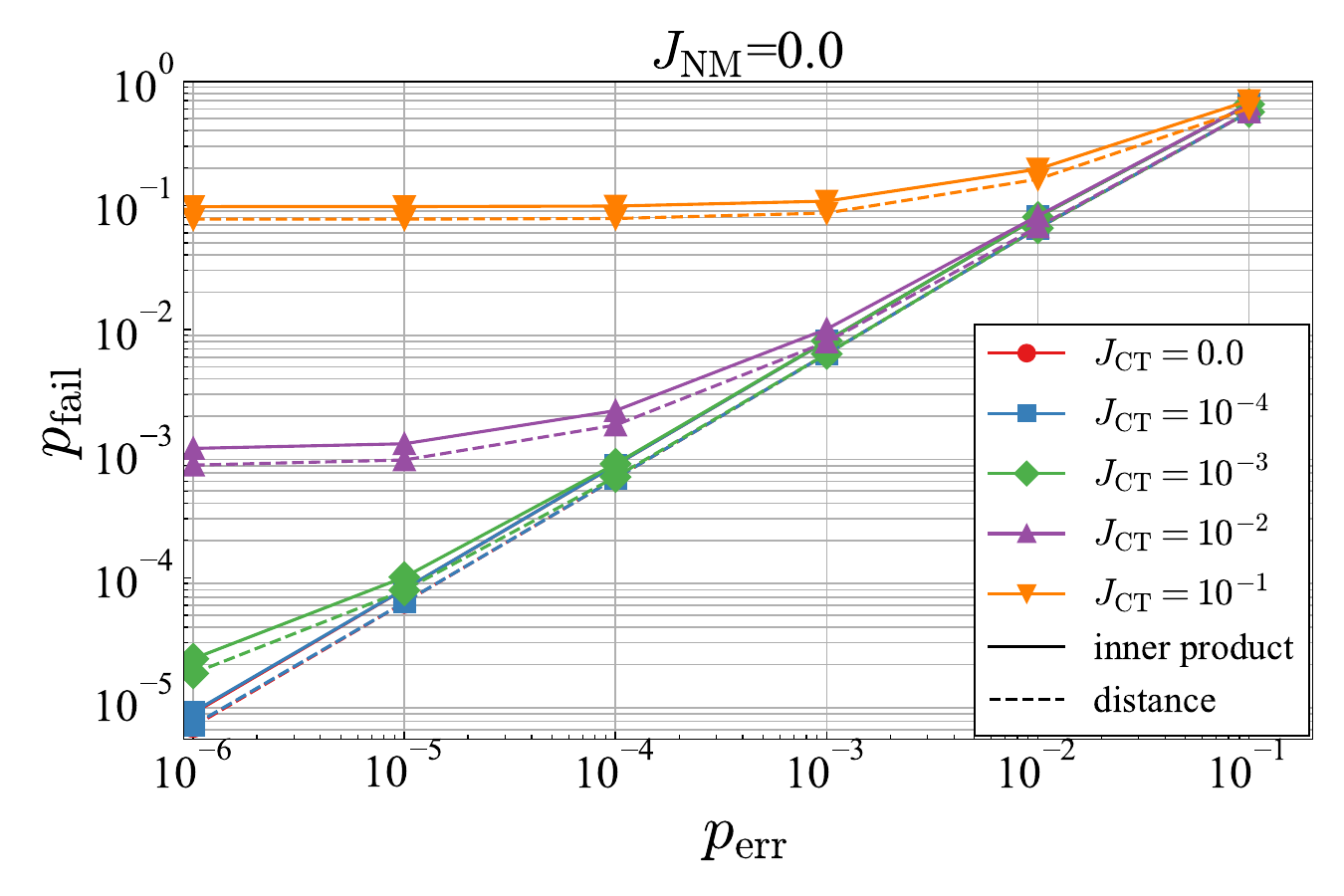}
}\\
\subfloat[\label{fig:cross_talk_rfr} Effect of both NM and CT error]{%
  \includegraphics[width=0.67\linewidth]{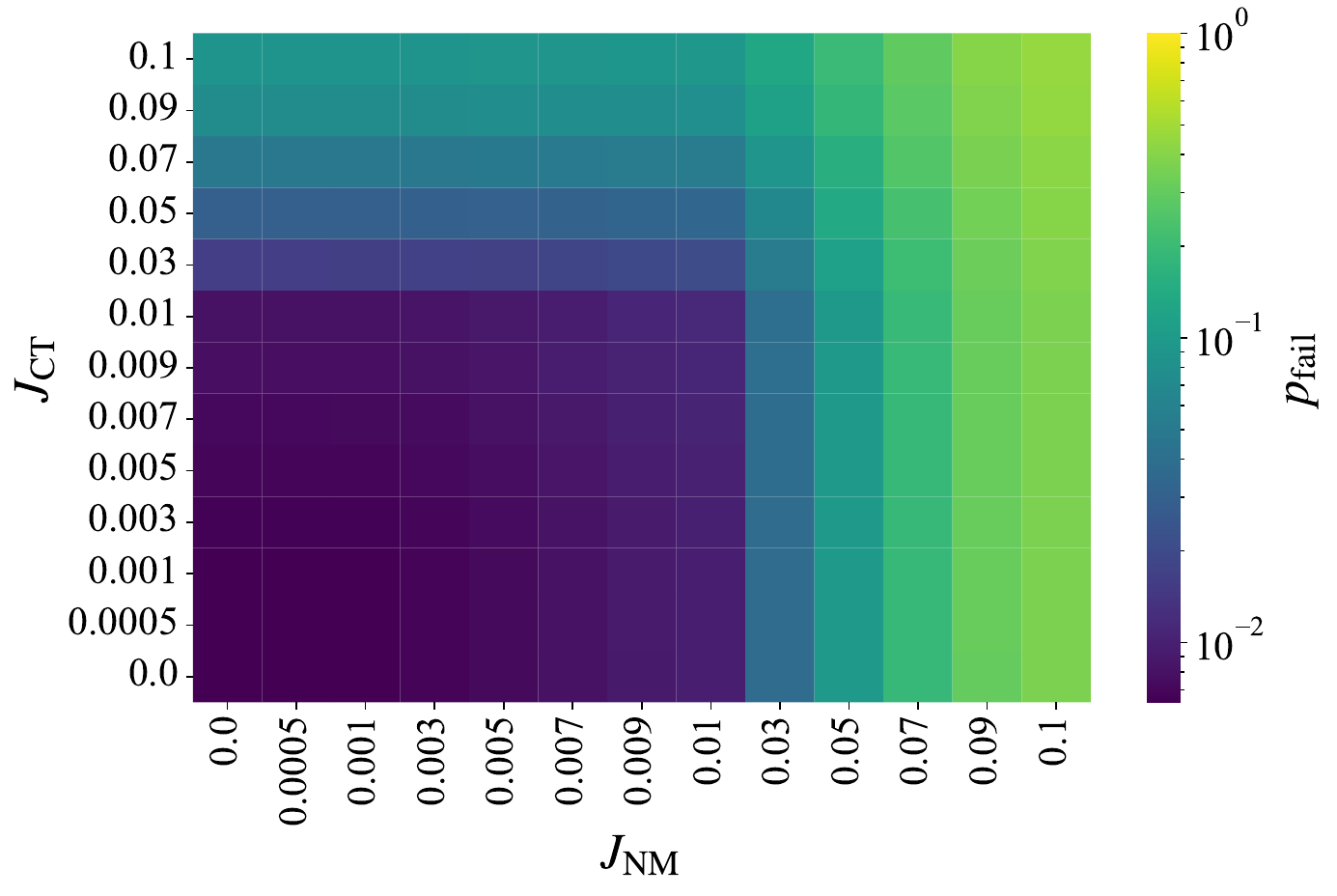}%
}
\caption{\label{fig:logical_failure_rate}
(a) Logical failure rate of 5-qubit error correction for the depolarizing error rate with only NM error and (b) Logical failure rate of 5-qubit error correction for the depolarizing error rate with only CT error. 
Coloured lines correspond to each intensity of CT or NM interaction $J_{\text{NM}}$ or $J_{\text{CT}}$.
The solid line and the dashed line represent two different metrics of the logical failure rate: the former uses the inner product metric \eqref{eq:ler_inner}, while the latter uses the channel distance \eqref{eq:ler_distance}.
Both plots indicate that stronger NM and CT interactions make the logical failure rate worse. However, a fair comparison must account for the added noise due to $J_{\text{NM}}>0$. The same care is not needed for $J_{\text{CT}}>0$ as cross-talk does not change the stochastic noise levels and only scrambles information.
(c) The dependency of logical failure rate $p_{\textrm{fail}}$ (using the inner product metric) for the CT and NM intensities under $p_{\textrm{err}}=10^{-3}$.}
\end{figure}

The logical failure rates we obtained are shown in Fig.~\ref{fig:logical_failure_rate}.
Panels (a) and (b) show the logical failure rate of the five-qubit code for a fixed depolarising error rate for the stochastic noise and varying rates of non-Markovian noise and crosstalk noise, respectively.
In Panel (b), the stochastic noise due to the depolarising channel is the same as when cross-talk is off. The cross-talk only scrambles the information on the data qubits. Nevertheless, this has an adverse effect on logical error rates.
Non-Markovianity, on the other hand, adds to the stochastic depolarising noise, as some information will inevitably be lost to the bath qubits. Thus, it should not be surprising that the logical error rates are higher in Panel (a) when compared to Panel (b) for the same coupling strength.
A fair analysis of the impact of non-Markovianity and cross-talk must account for the added noise due to the additional interactions with bath qubits and neighbouring qubits.

The logical failure rate is not suppressed even if $J_{\textrm{NM}}=J_{\textrm{CT}}=0$ on our strategy of maximum likelihood decoding.
This is because, as explained in the previous section, the error can avoid detection by the syndrome measurement if errors that a syndrome measurement captures at $k$ step occur after the syndrome measurement at $k$ step.
Even with a small error probability, the syndrome outcomes are scrambled, and the decoder cannot select the proper recovery operation. The third plot shows the logical failure rate with all non-Markovian noise, crosstalk noise and stochastic noise, while the error probability of stochastic noise is fixed $p_{\textrm{err}} = 1.0\times 10^{-3}$.
This heatmap shows that the effects of NM noise and CT noise suddenly appear at the $10$ times greater strengths of $J_{\textrm{CT}}$ and $J_{\textrm{NM}}$ than the error probability of stochastic noise.

The lack of performance from the QEC is not surprising. This five-qubit code is not designed to handle complex noise. Our goal is to display how QEC interacts with the process tensor, which we hope will lead to more sophisticated code design to overcome complex noise. 

The contraction of the large tensor, such as Fig.~\ref{fig:Whole_process} and corresponding to Eq.~\eqref{eq:chi_contraction}, is computationally challenging for the larger QEC codes. 
%five-qubit code. Representing this tensor already requires memories of $10+10$ qubit and $4$ classical bit.
To make our method scalable, we introduce the approximation method of tensor networks in the following section.

\begin{table}
    \caption{The stabiliser generators, logical operators and the pure errors of the five-qubit code.}\label{tab:five_qubit_code}
    \centering
    \begin{tabular}{c|c|c}
        Stabilizer generator & Logical operators & Pure errors\\ 
        \hline
        $G_1=X_1Z_2Z_3X_4$ & $X_L = X_1X_2X_3X_4X_5$ & $P_1=X_2$\\
        $G_2=X_2Z_3Z_4X_5$ & $Z_L = Z_1Z_2Z_3Z_4Z_5$ & $P_2=Z_5$\\
        $G_3=X_1X_3Z_4Z_5$ &  & $P_3=Z_3$\\
        $G_4=Z_1X_2X_4Z_5$ &  & $P_4=X_1$\\
    \end{tabular}
\end{table}

\subsection{Steane code and Approximation Methods}

We propose a method for scalable calculation of the logical failure rate and optimal decoding by approximating the process tensor and tester. While several approximation methods exist for contracting tensor networks, we employ a method based on Matrix Product States (MPS) in this study.

\subsubsection{MPS method for approximating process tensor and tester}

\begin{figure}[tbp]
    \centering
    \includegraphics[width=\linewidth]{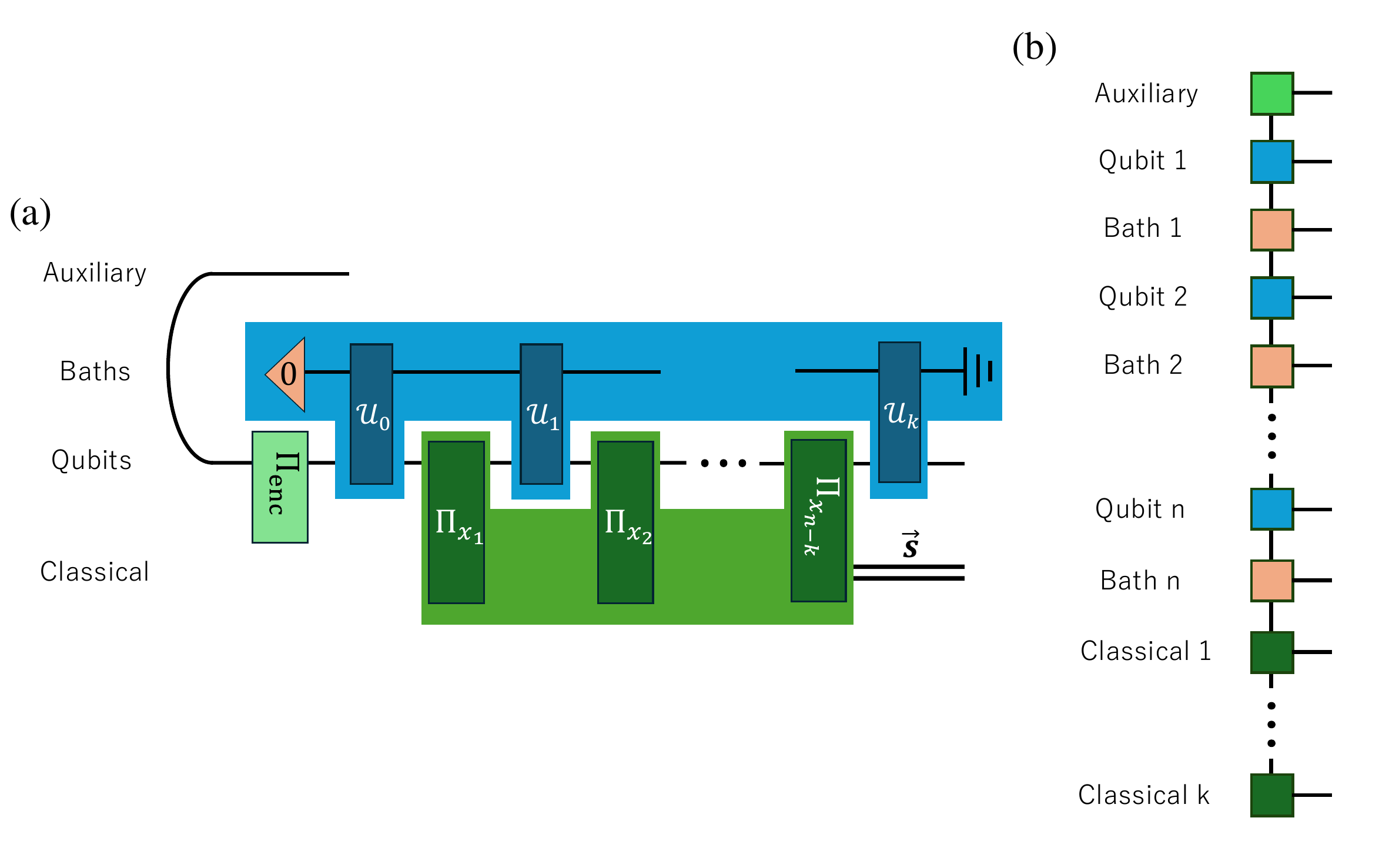}
    \caption{(a) The tensor network representing $\Upsilon[\Pi_{\mathrm{enc}}\otimes\Pi_{\outcomes{n-k}}]$. (b) The MPS structure for the state of the system.
    \label{fig:MPS}}
\end{figure}

The tensor network of $\chi_{\textrm{HS}}(L, \vec{s})$ in Fig.~\ref{fig:chi_TN} can be seen as contracting $\Upsilon[\choiprojector_{\outcomes{\timesteps{n-k}}}]\star\choiprojector_{\mathrm{enc}}$, as shown in Fig.~\ref{fig:MPS}(a), and a tensor network involving the decoder and perfect syndrome measurements.
Let us denote the latter tensor to
\begin{equation}
    \mathbf{Q}(L, \vec{s}) = \Pi_{\mathrm{dec}} \star\mathbf{I}_{\timesteps{n-k+1:2(n-k+1)+1}}[\mathbf{C}_{L_{\text{pdc}},\vec{s}_{\text{pdc}}}]\star\mathbf{R}(L, \vec{s}).
\end{equation}
$\mathbf{Q}(L, \vec{s})$ acts as the tensor object to assess the quality of the logical state and does not include the bath system. So it is significantly smaller than $\Upsilon[\choiprojector_{\outcomes{\timesteps{n-k}}}]\star\choiprojector_{\mathrm{enc}}$. Furthermore, $\mathbf{Q}(L, \vec{s})$ does not depend on the noise models and is prepared by only a single execution for each QEC code. Therefore, in this study, it is assumed that the computational cost of contracting $\mathbf{Q}(L, \vec{s})$ is sufficiently small compared to that of computing the former one, and the discussion focuses on the methods for approximating the calculation of $\Upsilon[\choiprojector_{\outcomes{\timesteps{n-k}}}]\star\choiprojector_{\mathrm{enc}}$.

Consider calculating the process tensor and tester using a state-vector simulator. In the model we are considering, the data system and the bath system consist of $n$-qubit systems, and the classical system consists of $k$-bit systems. Furthermore, a $1$-qubit auxiliary system is required to describe the quantum channel on $\mathscr{H}_{\textrm{logical}}$ through the Choi-Jamiolkowski correspondence. Simulating this process using a state-vector method requires a memory footprint of $4^{2n+1}\times2^k$. In the case of the Steane code where $n=7$ and $k=6$, the space complexity scales as $2^{36}$, necessitating parallel processing on large-scale high-performance computing to perform this simulation. 

In this study, instead, we consider representing the states in the process tensor and tester using MPS~\cite{perez-garcia_Matrix_2007a}. We plot the structure of MPS in Fig.~\ref{fig:MPS}(b). We first place the indices of the auxiliary system, followed by the alternating indices of the data system and the bath system, and finally, the indices of classical bits are aligned. We start by representing the initial state of $\doubleket{\Pi_{\mathrm{enc}}}$ using MPS. Noise and stabilizer measurements are all described as Matrix Product Operator (MPO), and time evolutions are performed using the MPO-MPS method. During this process, a maximum bond dimension of the MPS is set to $D$, and excess bond dimensions are truncated by using the canonical form~\cite{schollwoeck_densitymatrix_2011}. We also truncate small singular values below some threshold for further reduction in bond dimension. This procedure is repeated, and finally, the bath qubits are traced out to compute the desired MPS representation of the process tensor and tester.

\begin{figure*}
    \centering
    \includegraphics[width=1\linewidth]{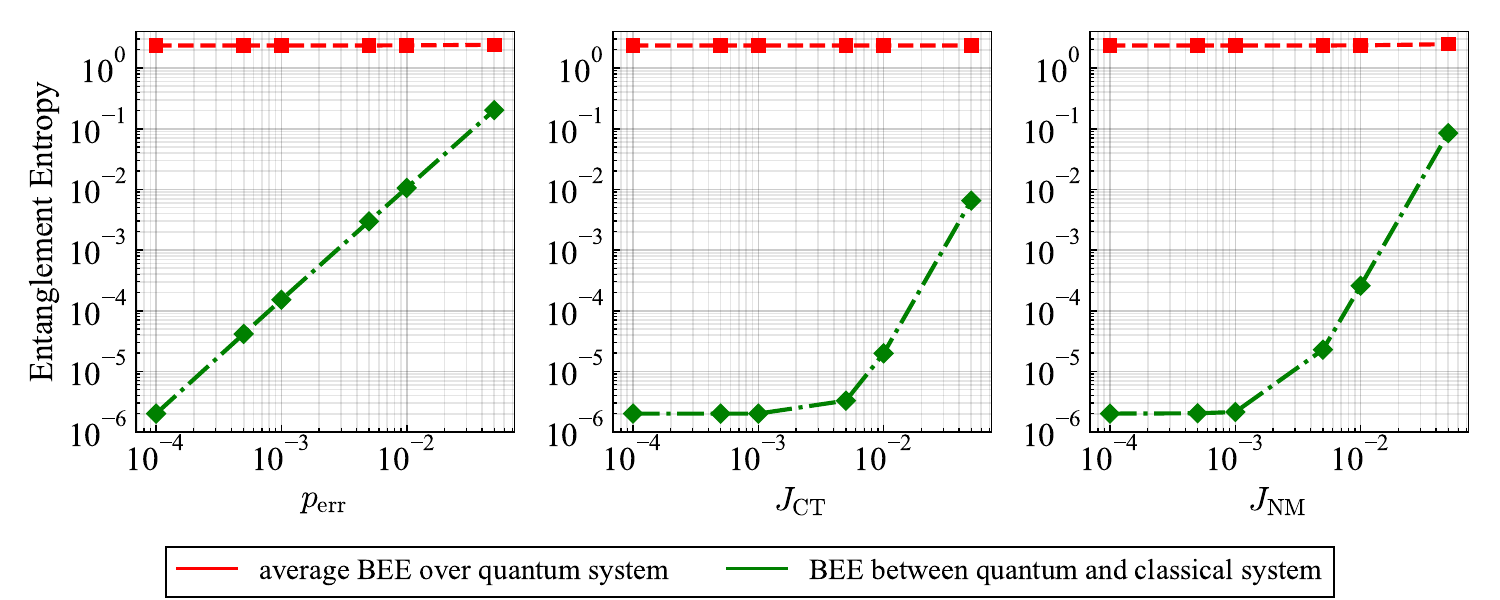}
    \caption{Bipartite entanglement entropy (BEE) of the final state of the five qubit code is plotted as a function of various noise parameters. All other values of noise parameters are fixed as $10^{-4}$ for each subplot. The red line represents the averaged BEE over quantum systems, and the green line represents the BEE between quantum and classical systems.
    \label{fig:ee_between_q_and_c}}
\end{figure*}

The accuracy of the MPS approximation strongly depends on the ordering of the indices. In this noise model, since each data qubit interacts with one bath qubit, we arrange the data and bath qubits alternately to correspond to this interaction and to improve the accuracy.
Furthermore, we positioned the indices corresponding to classical bits at the bottom side of the MPS. Fig.~\ref{fig:ee_between_q_and_c} illustrates the average of the bipartite entanglement entropy (BEE) of the final state of the 5-qubit code over the quantum system and the BEE between classical bits and the quantum system for each strength of noises.
Here, we treat the entire tensor, including classical systems, as a wave function and calculate the BEE between the classical system and the quantum system in a standard way.
Fig.~\ref{fig:ee_between_q_and_c} shows that, when the noise affecting the system is low, the correlation between the quantum system and classical system is much smaller than that of the quantum system. In cases of low noise, the patterns of emerged syndrome outcomes are biased, enabling the MPS algorithm to automatically truncate extremely unlikely syndrome patterns. Therefore, it is expected that adding classical bits to the end of the MPS will require relatively low additional computational resources.

One advantage of using the MPS method for process tensors is that it allows a scalable execution with respect to the number of qubits or stabilizer measurements by setting an upper bound on the bond dimension. Especially, an accurate approximation is possible when the QEC codes have a 1-D structure or the noise strength is relatively small.
On the other hand, there is no guarantee of representing the system well with an MPS, especially in the case of an increased number of qubits, multiple measurements, or strong noise. 
The initial state of the logical Bell pair, which is a superposition state of equal-weighted product states, requires accurate representation; if not, it significantly degrades precision.

\subsubsection{Numerical results}

There are two evaluation criteria for this method. The first is the accuracy of estimating the logical failure rate of stabilizer codes under noise. 
Classical simulations are used to investigate the performance of QEC codes under specific noise conditions before implementing them in actual devices.
Therefore, it is necessary that the approximations are sufficiently accurate to correctly capture the trends in logical failure rates. In that sense, the estimation of the logical failure rate with $2$-norm metric can be formulated as:
\begin{equation}
    p_{\mathrm{est}}=\sum_{\vec{s}\in\mathbf{s}} \tilde{p}(\vec{s})\min_L\|\mathbf{Q}(L, \vec{s})\star \tilde{\Upsilon}[\choiprojector_{\outcomes{\timesteps{n-k}}}(\vec{s})]\star\choiprojector_{\mathrm{enc}}/\tilde{p}(\vec{s}) - I\|_2
\end{equation}
where $\tilde{\Upsilon}, \tilde{p}$ are calculated with MPS approximation, and $\choiprojector_{\outcomes{\timesteps{n-k}}}(\vec{s})$ represents the projectors corresponding to the specific syndrome outcomes $\vec{s}$.

The second criterion is the performance evaluation when used as a tensor network decoder. The performance of this method as a TN decoder can be evaluated based on the following logical failure rate:
\begin{equation}
    p_{\mathrm{perf}}=\sum_{\vec{s}\in\mathbf{s}} p(\vec{s})\|\mathbf{Q}(\tilde{L}(\vec{s}), \vec{s})\star\Upsilon[\choiprojector_{\outcomes{\timesteps{n-k}}}(\vec{s})]\star \choiprojector_{\mathrm{enc}}/p(\vec{s}) - I\|_2
\end{equation}
where $\tilde{L}(\vec{s})$ is obtained via
\begin{equation}
    \tilde{L}(\vec{s})=\mathrm{argmin}_L\|\mathbf{Q}(L, \vec{s})\star \tilde{\Upsilon}[\choiprojector_{\outcomes{\timesteps{n-k}}}(\vec{s})]\star \choiprojector_{\mathrm{enc}}/\tilde{p}(\vec{s}) - I\|_2.
\end{equation}

\begin{figure*}
    \centering
    \includegraphics[width=1\linewidth]{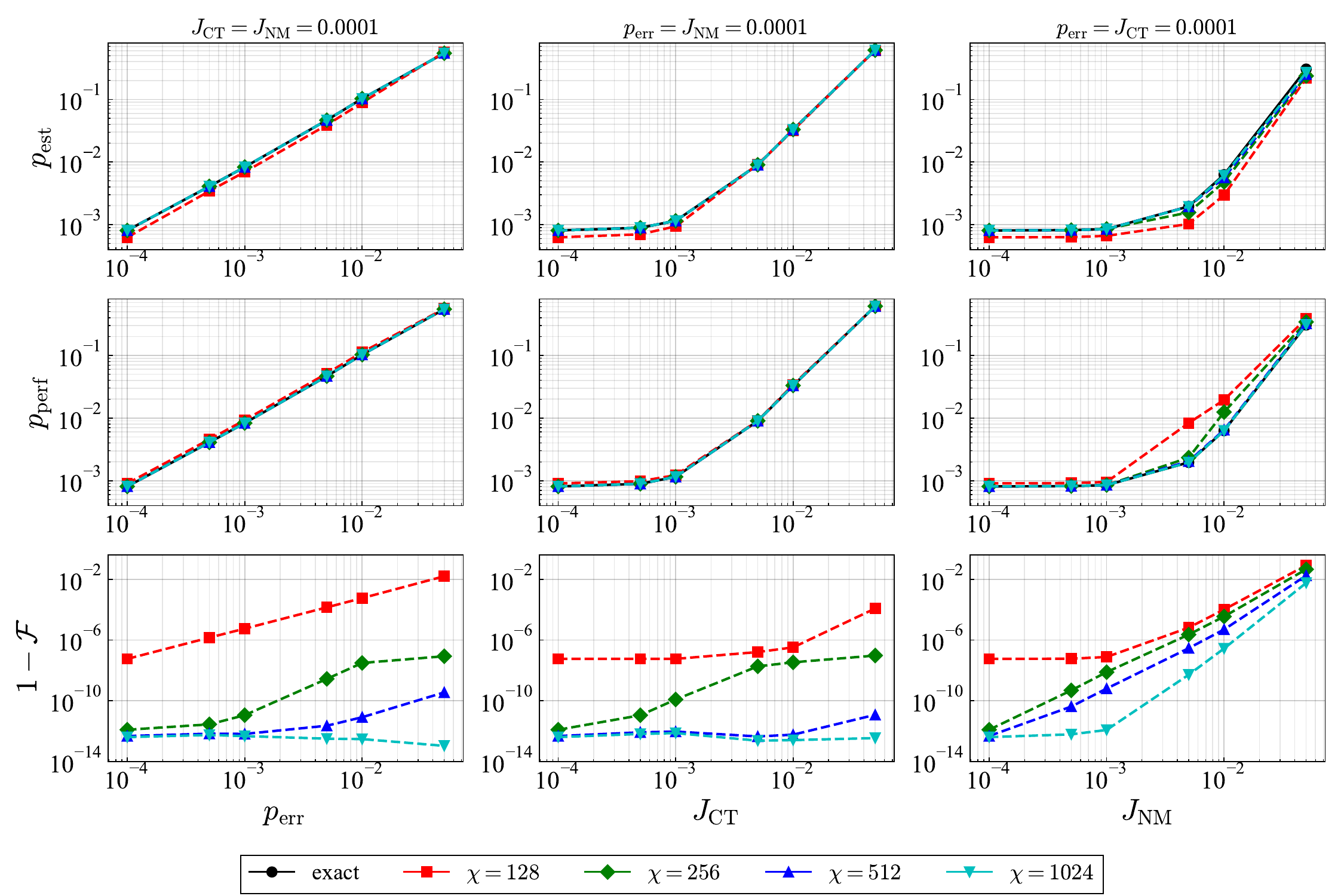}
    \caption{The estimation of the logical failure rate $p_\mathrm{est}$, the logical failure rate of approximated TN decoder $p_\mathrm{perf}$ and infidelity $1-\mathcal{F}$ of the Steane code is plotted as a function of noise strength $p_{\mathrm dep}, J_{\mathrm{CT}}$ and $J_{\mathrm{NM}}$ by fixing the other noise parameters. The black line represents the exact result from the tensor network contraction without approximations. The results of different maximum bond dimensions $\chi$ for the MPS approximation are plotted in different colours.}
    \label{fig:mps_result}
\end{figure*}

We executed numerical experiments with the Steane code, where state vector simulation is challenging, but exact TN contraction, as described in Sec.~\ref{sec:five-qubit}, is still possible. We set the threshold to be $10^{-8}$ and cut all singular values below this value during the canonical update.

Fig.~\ref{fig:mps_result} shows the estimates of the logical failure rate $p_{\mathrm{est}}$, the performance as a TN decoder $p_{\mathrm{perf}}$, and predicted fidelity values $\mathcal{F}$ obtained from MPS. 
First, we consider the relationship between fidelity, noise strength, and maximum bond dimension $\chi$. As shown in the lower panel, increasing noise parameters leads to a reduction in fidelity. This decrease can be attributed to the added complexity in the quantum system due to noise, as well as the increased correlation between the quantum and classical systems, as shown in Fig.~\ref{fig:ee_between_q_and_c}. Increasing the bond dimension can systematically improve fidelity.

Next, we investigate the relationship between fidelity and $p_{\mathrm{est}}, p_{\mathrm{perf}}$. In settings where fidelity is low, the estimated logical failure rate tends to deviate from the actual value. From the perspective of TN decoder performance, low fidelity does not necessarily imply performance degradation. In regions where depolarizing noise or crosstalk noise is strong, no significant performance deterioration is observed. However, in regions where non-Markovian noise is strong, a clear performance decrease is evident. In either case, preparing a larger bond dimension is effective in improving prediction accuracy and performance.

\begin{table*}[h!]
\centering
\begin{subtable}[t]{\textwidth}
\centering
\begin{tabular}{|c||c|c|c|c|}
\hline
 $\chi$ & $p_{\rm est}$ & $p_{\rm perf}$ & $1-\mathcal{F}$ & time[s] \\
\hline\hline
$128$ & $6.2682\times 10^{-4}$ & $9.0272\times 10^{-4}$ & $5.7262\times 10^{-8}$ & $6.3572 \times 10^{1}$ \\
\hline
$256$ & $8.0750 \times 10^{-4}$ & $8.0847 \times 10^{-4}$ & $1.2322\times 10^{-12}$ & $1.0816\times 10^{2}$ \\
\hline
$512$ & $8.0791 \times 10^{-4}$ & $8.0845 \times 10^{-4}$ & $4.9616 \times 10^{-13}$ & $1.8075 \times 10^{2}$ \\
\hline
$1024$ & $8.0794 \times 10^{-4}$ & $8.0845 \times 10^{-4}$ & $4.1056 \times 10^{-13}$ & $1.9941 \times 10^{2}$ \\
\hline
exact & $8.0828 \times 10^{-4}$ & $8.0828 \times 10^{-4}$ & 0.0 & $5.3214 \times 10^{2}$ \\
\hline
\end{tabular}
\caption{$p_{\mathrm{err}}=J_{\mathrm{CT}}=J_{\mathrm{NM}}=0.0001$}
\end{subtable}
\\[1em]
\begin{subtable}[t]{\textwidth}
\centering
\begin{tabular}{|c|c|c|c|c|}
\hline
 $\chi$  & $p_{\rm est}$ & $p_{\rm perf}$ & $1-\mathcal{F}$ & time[s] \\
\hline
$128$ & $6.7378 \times 10^{-3}$ & $9.6312 \times 10^{-3}$ & $5.7948 \times 10^{-6}$ & $1.7937 \times 10^{2}$ \\
\hline
$256$ & $8.6344 \times 10^{-3}$ & $8.6946 \times 10^{-3}$ & $9.8323 \times 10^{-9}$ & $4.2863 \times 10^{2}$ \\
\hline
$512$ & $8.6890 \times 10^{-3}$ & $8.6921 \times 10^{-3}$ & $1.2113 \times 10^{-9}$ & $8.7629 \times 10^{2}$ \\
\hline
$1024$ & $8.6908 \times 10^{-3}$ & $8.6913 \times 10^{-3}$ & $1.0829 \times 10^{-12} $& $1.3200 \times 10^{3}$ \\
\hline
exact & $8.6913 \times 10^{-3}$ & $8.6913 \times 10^{-3}$ & 0.0 & $5.3214 \times 10^{2}$ \\
\hline
\end{tabular}
\caption{$p_{\mathrm{err}}=J_{\mathrm{CT}}=J_{\mathrm{NM}}=0.001$}
\end{subtable}
\\[1em]
\begin{subtable}[t]{\textwidth}
\centering
\begin{tabular}{|c|c|c|c|c|}
\hline
 $\chi$ & $p_{\rm est}$ & $p_{\rm perf}$ & $1-\mathcal{F}$ & time[s] \\
\hline
$128$ & $1.1548 \times 10^{-1}$ & $1.4335 \times 10^{-1}$ & $8.5472 \times 10^{-4}$ & $2.7175 \times 10^{2}$ \\
\hline
$256$ & $1.3197 \times 10^{-1}$ & $1.3565 \times 10^{-1}$ & $1.0217 \times 10^{-4}$ & $6.6340 \times 10^{2}$ \\
\hline
$512$ & $1.3491 \times 10^{-1}$ & $1.3560 \times 10^{-1}$ & $1.3690 \times 10^{-5}$ & $2.1405 \times 10^{3}$ \\
\hline
$1024$ & $1.3509 \times 10^{-1}$ & $1.3549 \times 10^{-1}$ & $5.6104 \times 10^{-7}$ & $6.3783 \times 10^{3}$ \\
\hline
exact & $1.3549 \times 10^{-1}$ & $1.3549 \times 10^{-1}$ & 0.0 & $5.3214 \times 10^{2}$ \\
\hline
\end{tabular}
\caption{$p_{\mathrm{err}}=J_{\mathrm{CT}}=J_{\mathrm{NM}}=0.01$}
\end{subtable}
\\[1em]
\caption{Table illustrating the values of the estimation of the logical failure rate $p_\mathrm{est}$, the performance as a TN decoder $p_\mathrm{perf}$, fidelity $\mathcal{F}$ and the computing time of the Steane code. Numerical experiments are performed under (a) low, (b) medium, and (c) high noise conditions. The column $\chi$ represents the maximum bond dimension for the MPS methods and "exact" refers to the result from the exact tensor network contraction.}
\label{table:mps_time}
\end{table*}

We summarized the results for the logical failure rates and elapsed times using the MPS and TN methods across some noise parameter settings in Table.~\ref{table:mps_time}. 
Note that we dynamically truncate small singular values below the threshold during the process. As a result, even with the same maximum bond dimension, execution time can vary depending on the details of the system.
In regions with low noise, the MPS method achieves accurate estimation and high decoding performance nearly equivalent to exact methods with moderate bond dimensions, while requiring significantly less computation time than exact contraction. Conversely, in regions with high noise, the simulation with large bond dimensions can result in longer computation times than the exact methods. These results suggest that there is a trade-off between computation time and accuracy. Since the required accuracy of the estimation value and the performance of the decoder vary depending on the intended use, it is necessary to appropriately switch methods and maximum bond dimensions in response to these needs and details of the system.

\section{Conclusion}

In this paper, we formulated a new method to simulate the QEC with the process tensor and a method to obtain the maximum likelihood decoder.
We also evaluated the performance of the five-qubit code with the process tensor and the ML decoder.
Our formulation allows the ML decoder and its logical failure rate to be calculated by the single-shot contraction of the tensor network that combines the process tensor and the tester of syndrome measurements.
The performance of the five qubit code on our framework was not distinguished.
On the other hand, the effect of non-Markovian noise and crosstalk noise on the logical failure rate shows that the complex noise profiles may lead to a greater likelihood of logical errors. However, a careful analysis is needed to make such concrete statements.

Additionally, we proposed an approximation method to calculate the estimation of the logical failure rate and the ML decoder using MPS. A well-known method for simulating open quantum dynamics by approximating the process tensor with a tensor network is the Process Tensor in Matrix Product Operator (PT-MPO) form method~\cite{strathearn_Efficient_2018,jorgensen_Exploiting_2019,lacroix_MPSDynamicsjl_2024,cygorek_ACE_2024,fux_OQuPy_2024}. This method applies MPO approximations by exploiting the weak temporal correlations of the process tensor. However, in our case, the local Hilbert dimensions of the memory systems are already very large, making the PT-MPO method unsuitable. Instead, we represent both the process tensor and tester using MPS, avoiding the issue of exploding local Hilbert dimensions. Furthermore, by leveraging the weak correlations between the classical and quantum systems, we found that very fast and accurate simulation and decoding are possible, particularly in low-noise regions.

Our paper opens practical pathways for considering QEC with complex hardware noise. The future work will be to concrete the better construction of the ML decoder.
Our construction of the ML decoder treats just single round syndrome measurement. 
Imagine the decoding of the surface code with noisy measurements or circuit-based decoding, the decoder must be able to handle multiple rounds of syndrome measurements, and the parity of each temporal set of syndrome outcomes helps effectively on minimum perfect matching decoding~\cite{Dennis2002-iz}.
Considering the analogy of the minimum perfect matching decoding, combining syndrome outcomes for several rounds may be able to suppress the logical failure rate. Moreover, when considering multiple measurement rounds or larger QEC codes, the size of the tensor network representing the ML decoder becomes more enormous. In such a case, it may be necessary to employ higher-dimensional tensor network ansatz or apply general approximation methods for tensor network contraction such as compression-based methods~\cite{gray_Hyperoptimized_2022} or belief-propagation algoirithms~\cite{alkabetz_Tensor_2021,tindall_Gauging_2023a}.

Finally, this work will be useful as a metric to evaluate the efficiency of the quantum error correction code.
Tensor network code~\cite{Farrelly2021-yx, Cao2022-zw, Farrelly2022-xz} is a new framework to build the QEC code from seed tensor.
There are several challenges in optimising the blueprint of the Tensor-network code, such as contraction patterns~\cite{Su2023-wb, Mauron2024-oz}.
The logical failure rate of the code is the most important metric to evaluate the performance of the code.
Our method will be useful for evaluating the logical failure rate of the tensor network code and will be a powerful tool for optimising its blueprint.
In this aspect, our method paves the way for the new bottom-up construction of the quantum error correction code.

\section*{Acknowledgement}
F.K. was supported by JST Moonshot R\&D, Grant No. JPMJMS2066, ASPIRE, Grant No. JPMJAP2319, and the establishment of university fellowships towards the creation of science and technology innovation, Grant No. JPMJFS2125. H. Manabe is supported by JSTCOI-NEXT program Grant Numbers JPMJPF2014. G.A.L.W. is supported by an Alexander von Humboldt Foundation research fellowship.
K.M. is supported by the Australian Research Council under Discovery Projects DP210100597 and DP220101793.

\bibliographystyle{iop-num}
\bibliography{reference}

\end{document}